\documentclass[11pt, a4paper]{article} 
\usepackage[utf8]{inputenc}
\usepackage[british]{babel}
\usepackage[nottoc]{tocbibind}
\usepackage{authblk}

\usepackage{blindtext}
\usepackage[top=1in, bottom=1in,left=1in,right=1in]{geometry}
\usepackage{amssymb}
\usepackage{amsmath}
\usepackage{mathtools}
\usepackage{multicol}
\usepackage{tikz}

\usepackage{amsfonts}
\usepackage{graphicx}
\usepackage{subfig}
\usepackage{caption}
\usepackage{braket}
\usepackage{array}
\usepackage{float}
\usepackage{hyperref}
\usepackage{cite}
\newcolumntype{M}[1]{>{\centering\arraybackslash}m{#1}}
\newcolumntype{N}{@{}m{0pt}@{}}
\usepackage{enumerate}
\usepackage{bm}
\usepackage{lipsum}
\usepackage{xcolor}
\usepackage{mathptmx}  
\usepackage{mathrsfs}
\providecommand{\keywords}[1]
{
  \small	
  \textbf{\textit{Keywords---}} #1
}

\title{Entanglement dynamics in intensity-dependent double Jaynes-Cummings model for squeezed coherent thermal states}

\begin{document}

\def\correspondingauthor{\footnote{Corresponding author's email: mandalkoushik1993@gmail.com}}
\author[]{Koushik Mandal \correspondingauthor{}}
\affil[]{\textit{Department of Physics, Indian Institute of Technology Madras, Chennai, India, 600036}}

\date{}

\maketitle

\begin{abstract}
In this paper, the entanglement dynamics of different subsystems such as atom-atom, atom-field and field-field with radiation field in squeezed coherent thermal states for the intensity-dependent double Jaynes-Cummings model (IDDJCM) and double Jaynes-Cummings model (DJCM) are investigated. The effects of squeezed and thermal photons on entanglement are examined, revealing their complementary roles in shaping the entanglement behavior in both  models. One of the main features of the double Jaynes-Cummings model is the observation of  entanglement sudden death  for every subsystem. The effects of various interactions such as Ising interaction, single photon exchange interaction and dipole-dipole interaction on entanglement dynamics are studied. The effects of detuning, Kerr-nonlinearity on the entanglement dynamics are investigated for every subsystem. It is noticed that proper choice of the interaction parameters associated with photon exchange interaction, dipole-dipole interaction, detuning and Kerr-nonlinearity effectively remove entanglement deaths from the dynamics. 
\end{abstract}


\section{Introduction}\label{sec1}
It is well known that the Jaynes-Cummings (J-C) model is one of the most successful models to describe the atom-field interaction and to study the dynamics of various quantum optical aspects. Entanglement \cite{RevModPhys.81.865} is one of the important quantities which can be investigated using the Jaynes-Cummings model. In recent years, entanglement has become  ubiquitous and necessary for quantum applications. It has become a resource for quantum teleportation \cite{PhysRevLett.70.1895}, super dense coding \cite{PhysRevLett.69.2881}, entanglement swapping, cryptography etc. In many quantum optical systems such as in ion traps \cite{PhysRevA.76.041801,PhysRevA.95.043813, PhysRevA.97.043806}, cavity quantum electrodynamics (CQED) \cite{PhysRevA.82.053832, rosadon, AHMADI2011820, niemczyk2010circuit, haroche2020cavity, fink2008climbing, PhysRevLett.105.173601, PhysRevA.92.023810}, circuit quantum electrodynamics \cite{Mamgain_2023} etc., the entanglement is studied. All these physical platforms involve nontrivial interactions between atoms and fields. Jaynes-Cummings model provides a platform to study entanglements in these systems.

One of the successful extensions of the usual Jaynes-Cummings model is the double Jaynes-Cummings model (DJCM), which was first introduced by Yonac et. al., \cite{Yönaç_2006, PhysRevLett.97.140403, eberly2, YU2006393, Yönaç_2007, yu2005evolution, eberly3}. This model can be used to study the properties of entanglement in a more extended quantum optical systems which are mentioned before. A striking feature of the DJCM is the sudden disappearance or death of entanglement for a certain duration. This phenomenon of disappearance of entanglement is termed as entanglement sudden death (ESD). After Yonac's work, there are studies on various systems with this model \cite{PhysRevA.76.042313, Pandit_2018,li2020entanglement, jakubczyk2017quantum}. In particular in Ref. \cite{li2020entanglement}, the authors have studied entanglement dynamics when atoms interact with fields in coherent and squeezed states. They have shown that the factors such as atomic spontaneous decay rate, cavity decay rate and detuning have significant effects on the entanglement sudden death. To study the effects of interaction, the authors in Ref.  \cite{Pandit_2018,PhysRevA.101.053805} considered cavities with Ising type and photon exchange interactions between the cavities. The notable features of entanglement decay, sudden rebirth and sudden death have been investigated in the presence of intrinsic decoherence in Ref. \cite{obada2018influence}. In another work, Laha \cite{laha2023dynamics} has  considered a double Jaynes-Cummings model to investigate the role of  a beam-splitter, dipole-dipole interaction as well as Ising type interactions. The photonic modes in this work are considered to be either the photonic vacuum state, or the coherent states or the thermal states and the entanglements investigated are the qubit-qubit and the oscillator-oscillator type.

Another important successful generalization of the standard Jaynes-Cummings model is the intensity-dependent Jaynes-Cummings model (IDJCM)  \cite{Faghihi_2013, Baghshahi_2014}  which Buck and Sukumar introduced, which is known as the Buck-Sukumar model. There are various studies done using this Hamiltonian. In \cite{buvzek1989light}, Buzek found that in the intensity-dependent coupling, Jaynes-Cummings model with the coherent field, light-squeezing exhibits periodic revivals. He also investigated the influence of the initial state of an atom on the squeezing of light in detail. In \cite{PhysRevA.39.3196}, Buzek showed that IDJCM interacting with the Holstein-Primakoff SU(1,1) coherent state, the revivals of  radiation squeezing are strictly periodical for any value of  initial squeezing. He also found an expression for the atomic inversion exhibiting  exact periodicity of the population revivals. The spectrum of emitted light by a single atom interacting with a single mode radiation field in an ideal cavity via the intensity-dependent coupling was also studied by Buzek\cite{VBuzek_1990}. In \cite{Naderi_2011}, Naderi has shown that in the absence of the rotating wave approximation (RWA), the J-C Hamiltonian can be transformed into an intensity-dependent Hamiltonian. They studied the effects of the counter-rotating terms which appear in the intensity-dependent Hamiltonian on atomic inversion, atomic dipole squeezing, atomic entropy squeezing, photon counting statistics, field entropy squeezing, etc. In another work \cite{naderi2005theoretical}, Naderi et. al., has provided a theoretical scheme for the generation of nonlinear coherent states in a micromaser under an intensity-dependent J-C model. Lo et. al., \cite{LO1999557} have investigated the eigenenergy spectrum of the $k$-photon intensity-dependent J-C model without rotating wave approximations. They show that for $k\geq 2$ the $k$-photon intensity dependent, J-C model without RWA does not have eigenstates in the Hilbert space spanned by the photon number states, i.e., the model becomes ill-defined. Ng et. al.,\cite{ng2000exact} also investigated the eigenenergy spectrum of the IDJCM without rotating-wave approximation. Their analysis shows that counter-rotating terms in the Hamiltonian dramatically change the RWA energy spectrum and that the non-RWA spectrum can be approximated by the RWA spectrum only in the range of a sufficiently small coupling constant. They also showed that IDJCM without RWA is well-defined, only when the coupling parameter is below a certain critical value.

After Yonac's work on double Jaynes-Cummings model and the entanglement sudden death, various systems with double Jaynes-Cummings model have been studied; however, there are very few studies on the intensity-dependent double Jaynes-Cummings model (IDDJCM). In \cite{qin2012entanglement} Xie Qin and Fang Mao-Fa have investigated the entanglement dynamics of IDDJCM with two different initial radiation fields. One is a coherent state and another is a squeezed vacuum state. Motivated by these works, in this paper, we study the entanglement dynamics for both double Jaynes-Cummings and intensity-dependent double Jaynes-Cummings models with the radiation fields in squeezed coherent thermal states (SCTS), and, with the atoms in a Bell state. 

The intensity-dependent double Jaynes–Cummings model provides a more realistic framework by incorporating nonlinear nature of atom–field coupling, reflecting practical scenarios such as those encountered in nonlinear optics, trapped ions, or micromaser dynamics. On the other hand, the use of squeezed coherent thermal states (SCTS) as the initial radiation field offers a richer and more physically relevant model, capturing both thermal noise, coherence, and nonclassical features like squeezing. This combination of nonlinear coupling and realistic field states opens up a deeper exploration of how this structured light interacts with matter and influences entanglement evolution. The motivation behind this study is to understand how these modifications affect entanglement robustness, entanglement sudden death phenomena, and the potential for entanglement engineering in practical quantum optical systems.

One of the main objectives of this work is to investigate the effects of squeezed photons and thermal photons on the entanglement in a coherent background for DJCM and IDDJCM. Studying the effects of different kinds of interactions on the entanglement dynamics in the system is also an objective. In this paper, the effects of diverse interactions such as single photon exchange interaction, Ising interaction and dipole-dipole interaction on entanglement are investigated. We also investigate the effects of detuning, Kerr-nonlinearity \cite{PhysRevA.45.6816, PhysRevA.45.5056, PhysRevA.44.4623, ahmed2009dynamics, sivakumar2004nonlinear, Mo_2022, PhysRevB.105.245310, zheng2017intrinsic, PhysRevA.93.023844, baghshahi2014entanglement, Faghihi:13, baghshahi2015generation} etc. Earlier, in Ref.\cite{mandal2023atomic}, the authors have studied the effects of the squeezed and thermal photons on entanglements in Jaynes-Cummings model. Here, it was shown how the tussling between ``classical noise'' (thermal photons) and ``quantum noise'' (squeezed photons) affects the quantum optical properties of the radiation field, atomic inversion and entanglement dynamics for Jaynes-Cummings model. The work presented here is a natural extension of our previous work to investigate the effects of the perpetual tussle between thermal and squeezed photons on the entanglement dynamics of various subsystems in DJCM and IDDJCM.

This work attempts to provide a comprehensive understanding of the interplay between interaction mechanisms, nonlinearity, and radiation field characteristics, with the objective to extract insights into controlling and enhancing entanglement in quantum optical systems. The results are of significance for applications in quantum information processing and the design of quantum communication protocols.

The paper is organized as follows: in section 2, the double Jaynes-Cummings (DJCM) and intensity-dependent double Jaynes-Cummings model (IDDJCM) are described. Section 3 shows the entanglement dynamics for DJCM and IDDJCM. The effects of photon exchange interaction are studied in section 4. Section 5 deals with the effects of dipole-dipole interaction. Next, the effects of Ising interaction is studied in section 6. In section 7, effects of Kerr-nonlinearity are investigated. The effects of detuning are dealt with in section 8.  Finally, in section 9, we conclude our results.

\section{The model}
\subsection{Photonic state}
Squeezed coherent thermal states (SCTS) are mixed states because of the of thermal photons. The density operator for SCTS is defined as\cite{PhysRevA.47.4474, PhysRevA.47.4487, yi1997squeezed}

\begin{equation}
\hat{\rho}_{\text{SCT}} = \hat{D}(\alpha)\hat{S}(\zeta)\hat{\rho}_{\text{th}}\hat{S}^{\dagger}(\zeta)\hat{D}^{\dagger}(\alpha),
\label{rho_scts}
\end{equation}
where
\begin{equation}
\hat{D}(\alpha) = \exp(\alpha \hat{a}^{\dagger} - \alpha^{*} \hat{a})
\end{equation}
is the displacement operator, for $\alpha$  a complex parameter; $\hat{a}$ and $\hat{a}^{\dagger}$ are the photon annihilation and creation operators respectively and
\begin{equation} 
\hat{S}(\zeta) = \exp\left(-\frac{1}{2}\zeta \hat{a}^{\dagger2} + \frac{1}{2} \zeta^{*}\hat{a}^{2}\right)
\end{equation}
is the squeezing operator with $\zeta = r e^{i\varphi}$; where $\zeta$ is the squeezing parameter;  $r$ and $\varphi$ denote the amplitude and phase of $\zeta$ respectively. The density operator of a thermal radiation field with a heat bath at temperature $T$ can be written as 
\begin{equation}
\hat{\rho}_{\text{th}} = \frac{1}{1 + \bar{n}_{th}}\sum_{n=0}^{\infty}\left(
\frac{\bar{n}_{th}}{\bar{n}_{th} + 1}\right)^{n}\ket{n}\bra{n},
\end{equation}
where $\bar{n}_{th}$ is the average number of thermal photons and it is given by
\begin{equation}
    \bar{n}_{th} = \frac{1}{\exp\left(\frac{h \nu}{k_B T}\right)-1};
    \label{avg_ther}
\end{equation}
$k_B$ is Boltzmann constant and $\nu$ is linear frequency of radiation field in Eq. (\ref{avg_ther}). 
The analytic expression for the PCD of SCTS can be written as\cite{PhysRevA.47.4474, PhysRevA.47.4487}

\begin{align}
P(l) =& \bra{l}\hat{\rho}_{\text{SCT}}\ket{l}\\
 =&~ \pi Q(0) \tilde{X}^{l}\sum_{q=0}^{l}\frac{1}{q!}\left(\frac{l}{q}\right)\Big|\frac{|\tilde{Y}|}{2 \tilde{X}}\Big|^{q}\nonumber\\
&\times \big|H_{q}((2Y)^{-1\slash 2} \tilde{Z})\big|^{2},
\end{align}
where $\pi Q(0) = R(0,0)$; $R$ is Glauber's $R$-function\cite{PhysRev.131.2766}; (see \ref{app_A} for more detail).

The average number of coherent photons is defined as 
\begin{equation}
    \bar{n}_c = |\alpha|^2,
\end{equation}
and the average number of squeezed photons is defined as
\begin{equation}
    \bar{n}_s = \sinh^2 r.
\end{equation}

\subsection{Atomic state}
The double Jaynes-Cummings model involves two two-level atoms: atom A and atom B, each one interacting with a cavity mode. Atom A is placed inside cavity a and atom B is placed inside cavity b. These two-level systems are initially prepared in either as a pure state or as a mixed state. For the pure state a maximally entangled Bell state is considered. In this paper, we use the entangled atomic state of the form
\begin{equation}
\ket{\psi_{\text{AB}}} = \cos \theta \ket{e, g} + \sin \theta \ket{g, e},
\label{bellstate}
\end{equation}
where \(\ket{e, g} = \ket{e} \otimes \ket{g}\) denotes the joint state in which atom A is in the excited state \(\ket{e}\) and atom B is in the ground state \(\ket{g}\), and vice versa for \(\ket{g, e}\).

\subsection{The Hamiltonian of the system}
Now, we introduce the Hamiltonian for IDDJCM  which is given by
\begin{align}
    \hat{H} &= \omega \hat{\sigma}_{z}^{\text{A}} + \omega \hat{\sigma}_{z}^{\text{B}} + \,  \nu \hat{a}^{\dagger} \hat{a} + \nu \hat{b}^{\dagger} \hat{b} + \lambda \left(\sqrt{\hat{N}_{\text{a}}}\,\hat{a}^{\dagger} \sigma_{-}^{\text{A}} + \hat{a}\sqrt{\hat{N}_{\text{a}}}\, \hat{\sigma}_{+}^{\text{A}}\right) \,\nonumber\\
& + \lambda \left(\sqrt{\hat{N}_{\text{b}}}\,\hat{b}^{\dagger} \hat{\sigma}_{-}^{\text{B}} + \hat{b}\sqrt{\hat{N}_{\text{b}}}\, \hat{\sigma}_{+}^{\text{B}}\right).
\label{iddjcm_hamil}
\end{align}
The factor $\hat{\sigma}_{z}^{i}$ represents the Pauli matrix in the $z$-basis and
$\hat{\sigma}_{+}^{i}$ and $\hat{\sigma}_{-}^{i}$ are the spin raising and lowering operators respectively. The index $i$ represents the atomic label. The photonic operators $\hat{a}$ and $\hat{b}$ are the annihilation operators corresponding to the two different cavities and the operators $\hat{a}^{\dagger}$ and $\hat{b}^{\dagger}$ are the corresponding creation operators. The coupling constant is represented by $\lambda$ and it describes the strength of the atom-field interaction with $\omega$ and $\nu$ being the atomic transition frequency and the radiation frequency respectively. $\hat{N}_{\text{a}}$ and $\hat{N}_{\text{b}}$ are the number operators for fields in cavities a and b respectively. For $\hat{N}_{\text{a}} = 1$ and $\hat{N}_{\text{b}} = 1$, we get the usual Hamiltonian for DJCM

\begin{equation}
    \hat{H'} = \omega \hat{\sigma}_{z}^{\text{A}} + \omega \hat{\sigma}_{z}^{\text{B}} + \, \nu \hat{a}^{\dagger} \hat{a} + \nu \hat{b}^{\dagger} \hat{b} + \lambda \left(\hat{a}^{\dagger} \sigma_{-}^{\text{A}} + \hat{a} \hat{\sigma}_{+}^{\text{A}}\right) 
+ \lambda \left(\,\hat{b}^{\dagger} \hat{\sigma}_{-}^{\text{B}} + \hat{b}\hat{\sigma}_{+}^{\text{B}}\right).
\label{ndjcm_hamil}
\end{equation}
Density operator for the atomic state is 
\begin{equation}
    \hat{\rho}_\textsubscript{AB}(0) = \ket{\psi_\textsubscript{AB}} \bra{\psi_\textsubscript{AB}},
    \label{rho_atom}
\end{equation} 
and initial density operator representing the field states is $\hat{\rho}_{\text{F}}(0)$. So, the density operator for the whole system can be written as 
\begin{align}
    \hat{\rho}_{\text{tot}}(0) =& \hat{\rho}_\textsubscript{AB}(0) \otimes \hat{\rho}_{\text{F}}(0).\\
    =& \hat{\rho}_\textsubscript{AB}(0) \otimes \hat{\rho}^{\text{a}}_{\text{F}}(0) \otimes \hat{\rho}^{\text{b}}_{\text{F}}(0),
    \label{rhotot_ini}
\end{align}
where $\hat{\rho}^{\text{a}}_{\text{F}}(0)$ and $\hat{\rho}^{\text{b}}_{\text{F}}(0)$ are the density operator for the radiation field in cavities a and b respectively. Since the fields in the cavities are not entangled initially, we can express $\hat{\rho}_{\text{F}}(0)$ as $\hat{\rho}^{\text{a}}_{\text{F}}(0) \otimes \hat{\rho}^{\text{b}}_{\text{F}}(0)$.
\subsection{Time evolution of the system}
To get the time evolve state of the system, we will use the interaction picture. The Hamiltonian in the interaction picture with the resonant condition (i.e., $\nu - \omega = 0$) can be written as 
\begin{equation}
    \hat{H}_{I} = \lambda \left(\sqrt{\hat{N}_{\text{a}}}\,\hat{a}^{\dagger} \sigma_{-}^{\text{A}} + \hat{a}\sqrt{\hat{N}_{\text{a}}}\, \hat{\sigma}_{+}^{\text{A}}\right) 
+ \lambda \left(\sqrt{\hat{N}_{\text{b}}}\,\hat{b}^{\dagger} \hat{\sigma}_{-}^{\text{B}} + \hat{b}\sqrt{\hat{N}_{\text{b}}}\, \hat{\sigma}_{+}^{\text{B}}\right),
\end{equation}
and similarly, it can be written for the Hamiltonian in Eq. (\ref{ndjcm_hamil}).

The time evolution operator for the whole system is given by
\begin{equation}
\hat{U}(t) = e^{\frac{-i\hat{H}_{I}t}{\hbar}},
\end{equation}
where $\hat{H}_{I}$ is the interaction Hamiltonian for the whole system. If there is no cross-interaction present in the system, this $\hat{U}(t)$ can be written as
\begin{equation}
    \hat{U}(t) = \hat{U}^{(1)}(t) \otimes \hat{U}^{(2)}(t),
\end{equation}
where
\begin{equation}
    \hat{U}^{(1)} (t) = \exp\left[\frac{ -i\lambda \left(\sqrt{\hat{N}_{\text{a}}}\,\hat{a}^{\dagger} \sigma_{-}^{\text{A}} + \hat{a}\sqrt{\hat{N}_{\text{a}}}\, \hat{\sigma}_{+}^{\text{A}}\right) t}{\hbar}\right], 
\end{equation}
and
\begin{equation}
    \hat{U}^{(2)} (t) = \exp\left[ \frac{-i \lambda \left(\sqrt{\hat{N}_{\text{b}}}\,\hat{b}^{\dagger} \hat{\sigma}_{-}^{\text{B}} + \hat{b}\sqrt{\hat{N}_{\text{b}}}\, \hat{\sigma}_{+}^{\text{B}}\right) t}{\hbar}\right]. 
\end{equation}

The initial state of the whole system $\hat{\rho}_{\text{tot}}(0)$ in Eq. \ref{rhotot_ini} can be written as 
\begin{equation}
    \hat{\rho}_{\text{tot}}(0) = \sum_{i,j} c_{ij} \, |i\rangle \langle j| \otimes \hat{\rho}^{\text{a}}_{\text{F}}(0) \otimes \hat{\rho}^{\text{b}}_{\text{F}}(0),
\end{equation}
where $c_{ij}$ are the coefficients containing different combinations of $\sin \theta, \cos \theta$. So, the time-evolved density operator for the whole system can be written as

\begin{equation}
\hat{\rho}_{\text{tot}}(t) = \hat{U}(t) \left(\sum_{i,j} c_{ij} \, |i\rangle \langle j| \otimes \hat{\rho}^{\text{a}}_{\text{F}}(0) \otimes \hat{\rho}^{\text{b}}_{\text{F}}(0)\right)\hat{U}^{\dagger}(t).
\label{rho_timeevo}
\end{equation}
This density operator can be expanded further in atomic basis. The details are shown in \ref{app_B}.\\

\noindent \textbf{{Computing entanglement for different subsystems:}}\\

Although, an analytical expression for the density operator corresponding to a squeezed coherent thermal state (SCTS) is highly involved and no closed-form expansion in the Fock basis is known, the entanglement dynamics of the system can still be reliably and robustly studied through numerical methods. In particular, by evolving the total density matrix and performing partial traces over the appropriate subsystems, one can investigate the entanglement between atoms, fields, and hybrid subsystems. For instance, to compute the atom-atom entanglement, we can trace out the field-field subsystems from the time evolved density operator in Eq. \ref{rho_timeevo}. The reduced atom-atom density operator can be written as:
\begin{align}
\hat{\rho}_{\textsubscript{AB}}(t) =& \text{Tr}_\text{a,b}[\hat{\rho}_{\text{tot}}(t)]\\
=& \sum_{i,j} \text{Tr}_{\textsubscript{a,b}}\left[ \hat{U}(t) \left( |i\rangle \langle j| \otimes \hat{\rho}^{\text{a}}_{\text{F}}(0) \otimes \hat{\rho}^{\text{b}}_{\text{F}}(0)\right)\hat{U}^{\dagger}(t)\right].
\end{align}
To compute the entanglement between atom A and field a, we trace out the atom B and field b degrees of freedom;
\begin{equation}
        \hat{\rho}_{\textsubscript {A,a}}(t) = \mathrm{Tr}_{\textsubscript{B,b}}[\hat{\rho}_{\text{tot}}(t)],
\end{equation}
and similarly for other subsystems.

\subsection{Entanglement measures}
To characterize the dynamics of entanglement, we need to measure the entanglement in the system. In this work, we investigate the dynamics of the bipartite entanglements like the atom-atom entanglement, and atom-field entanglement. The atom-atom entanglement can be conclusively measured using concurrence defined in \cite{wootters2001entanglement}
\begin{equation}
C_{\text{AB}} = \text{max}\{0, \Lambda_{1} - \Lambda_{2}-\Lambda_{3}-\Lambda_{4}\},
\end{equation}
where $\Lambda_{i} (i = 1, 2, 3, 4)$ are the decreasingly ordered square roots of the eigenvalues of the matrix 
$\hat{\rho} \left(\hat{\sigma}_{y}^{\text{A}}\otimes \hat{\sigma}_{y}^{\text{B}}\right) \hat{\rho}^{*}\\
\left(\hat{\sigma}_{y}^{\text{A}} \otimes \hat{\sigma}_{y}^{\text{B}}\right)$ and $\hat{\rho}$ is the two qubit atom-atom reduced density matrix. The value of concurrence lies in the range $0 \leq C \leq 1$, where $C=0$ implies a separable state and $C=1$ denotes a maximally entangled state. Though concurrence can be used to compute entanglement in both pure and mixed states, if the system is $2 \otimes 2$ dimensions. So, for higher dimensional systems we need to use other measures. In particular, when we consider the atom-field subsystem we are looking at a bipartite object, whose Hilbert space is composed of a 2-dimensional Hilbert space (associated with the atom) in a tensor product with a $\infty$-dimensional Hilbert space (associated with a continuous variable system i.e. the field. For these systems it is convenient to use the negativity \cite{wei2003maximal} which is defined as 
\begin{equation}
N(\rho)=\sum_{k}\Big( |\xi_{k}|-\xi_{k} \Big)/2, 
\end{equation}
where $\xi_{k}$ are the eigenvalues of $\hat{\rho}^{\text{PT}}$, the partial transpose of the density matrix, i.e., the matrix which is transposed with respect to any one of the subsystems. 
\begin{figure}[ht]
    \centering
    \includegraphics[scale = 0.45]{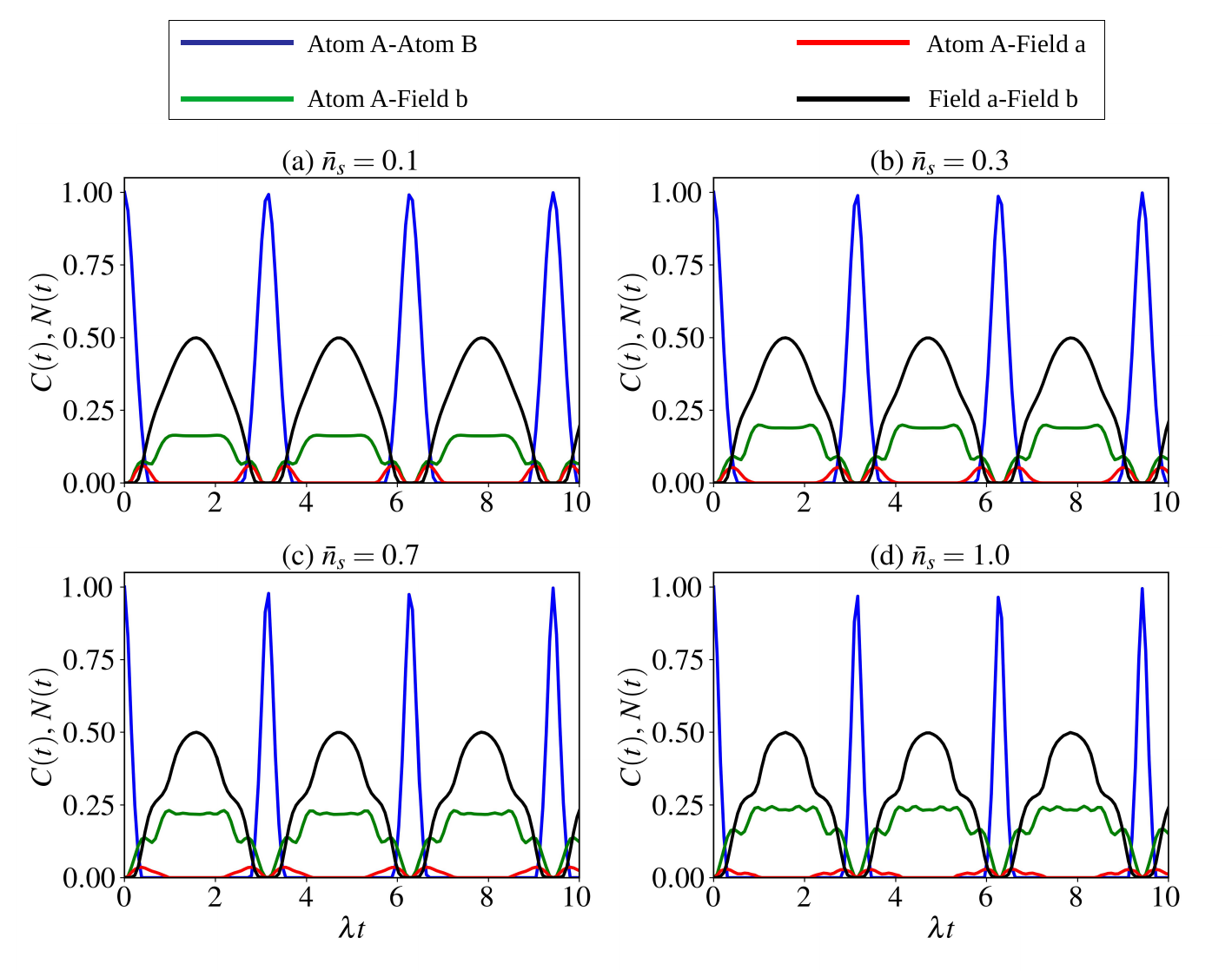}
    \caption{Entanglement dynamics for atom A-atom B, atom A-field a, atom a-field b and field a-field b with atoms in a Bell state and field in SCTS for IDDJCM. The values of the parameters used in these plots are $\bar{n}_c = 2$, $\bar{n}_{th} = 0.1$, $\bar{n}_s = 0.1, 0.3, 0.5, 1.0$ and $\theta = \frac{\pi}{4}$.}
    \label{fig_1}
\end{figure}

\begin{figure}[ht]
    \centering
    \includegraphics[scale = 0.35]{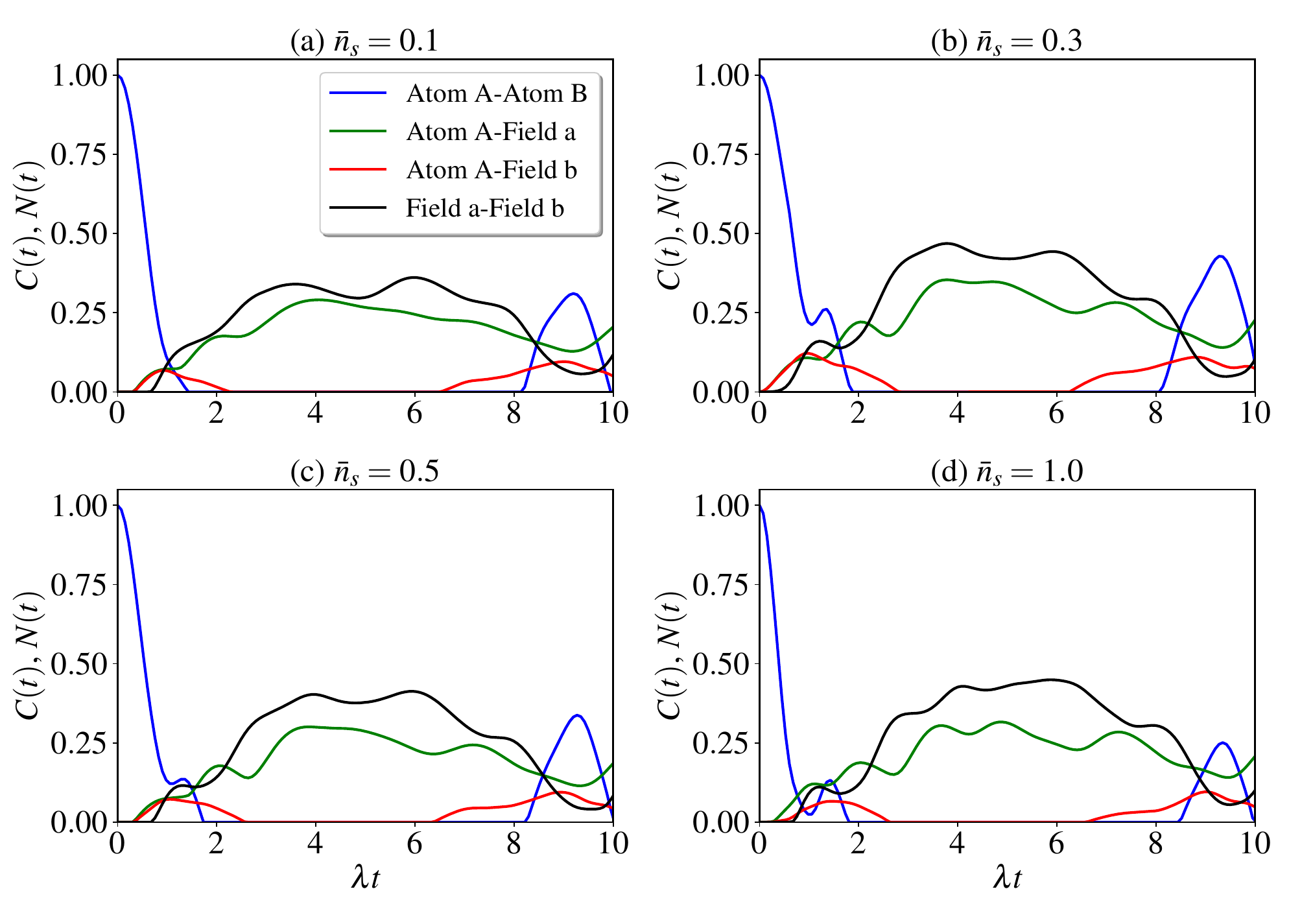}
    \caption{Entanglement dynamics for atom A-atom B, atom A-field a, atom a-field b and field a-field b with atoms in a Bell state and field in SCTS for DJCM. The values of the parameters used in these plots are $\bar{n}_c = 2$, $\bar{n}_{th} = 0.1$, $\bar{n}_s = 0.1, 0.3, 0.5, 1.0$ and $\theta = \frac{\pi}{4}$.}
    \label{fig_2}
\end{figure}
\section{Entanglement Dynamics in IDDJCM and DJCM with SCTS}

In this section, we investigate the entanglement dynamics of an atom-field system interacting with a squeezed coherent thermal state (SCTS) of the radiation field, where the atoms are initially prepared in a Bell state. Figure~\ref{fig_1} illustrates the entanglement dynamics for the intensity-dependent double Jaynes–Cummings model (IDDJCM). The parameters chosen for this study are $\bar{n}_c = 2$, $\bar{n}_{th} = 1.0$, and $\bar{n}_s = 0.1,\ 0.3,\ 0.5,\ 1.0$. Previous studies on the intensity-dependent Jaynes–Cummings model have established that the system dynamics exhibit a periodic structure~\cite{qin2012entanglement}. 

The chosen values \(\bar{n}_c = 2\), \(\bar{n}_{\text{th}} = 1\), and \(\bar{n}_s = 0.1\text{--}1.0\) ensure a physically meaningful balance between classicality and nonclassicality, while remaining experimentally realistic and numerically stable. This allows us to explore how thermal photons, coherent photons, and squeezing individually and collectively influence the entanglement dynamics in the IDDJCM framework.

From Fig.~\ref{fig_1}, it is evident that the atom-atom entanglement $C(t)$ (blue curve), as well as the negativity $N(t)$ for the three hybrid and field-field subsystems, all display periodic behavior (see \ref{app_C} for alternative figure). Specifically, the dynamics of $C(t)$ remain largely unchanged with increasing values of squeezed photons $\bar{n}_s$. The duration of the ESDs varies slightly with $\bar{n}_s$, but the peak amplitudes remain unaffected. A similar behavior is observed for the field-field subsystem (black curve in Fig.~\ref{fig_1}), where both the shape and amplitude of $N(t)$ are invariant under changes in $\bar{n}_s$. The explanation for this periodic behaviour of entanglement dynamics is as follows: in IDDJCM, the atom--field coupling strength becomes a function of the photon number, such as \( \lambda(n) \sim \sqrt{n} \) or a saturating form, rather than remaining constant as in the standard Jaynes--Cummings model. This intensity-dependent coupling plays a crucial role in stabilizing the entanglement dynamics. Specifically, it reduces the sensitivity of the system to fluctuations in the average photon number. In the standard model, an increase in field strength leads to proportionally faster Rabi oscillations, as the atom effectively experiences a stronger driving field. However, in the IDDJCM, the nonlinear or saturating nature of the coupling ``flattens'' this response, keeping the atom--field interaction effectively bounded even as the photon number increases.

As a result, characteristic features such as entanglement sudden deaths (ESDs) and the amplitude and timing of entanglement peaks remain relatively unchanged across a range of average photon numbers. This effect is particularly pronounced in subsystems like atom--atom and field--field, where the dynamics are dominated by coherent rephasing, and are less affected by photon number–induced dephasing. Thus, the intensity-dependent coupling provides a form of dynamical insulation against variations in field intensity, leading to more robust and predictable entanglement behavior.

Entanglement corresponding to the atom A–field $a$ subsystem (green curve), the amplitude of $N(t)$ increases with $\bar{n}_s$, while the overall shape of the dynamics and the ESD characteristics remain unchanged. Conversely, in for the atom A–field $b$ subsystem (red curve), both the peak amplitudes and the durations of the ESDs decrease as $\bar{n}_s$ increases. Overall, the increase in the number of squeezed photons does not significantly impact the entanglement dynamics of the field-field subsystem, and the curves for different $\bar{n}_s$ values largely overlap.

Figure~\ref{fig_2} presents the entanglement dynamics for the standard double Jaynes–Cummings model (DJCM). Unlike the IDDJCM, the entanglement in DJCM does not exhibit a periodic, pulse-like structure (see \ref{app_C} for alternative figure). In the case of atom-atom entanglement, $C(t)$ begins at its maximum value and then undergoes a sharp decline. A pronounced ESD occurs, and its duration increases with rising $\bar{n}_s$. The influence of thermal photons is clearly visible in this behavior. A significant distinction arises between the intensity-dependent double Jaynes–Cummings model (IDDJCM) and the standard double Jaynes–Cummings model (DJCM) in terms of entanglement behavior. In the standard DJCM, the Rabi frequency scales as $\Omega_n = 2 \lambda \sqrt{n + 1}$, leading to a broad spread of frequencies associated with different photon number states. This results in dephasing due to destructive interference, thereby suppressing entanglement revivals and producing aperiodic or damped dynamics.

In \cite{Mandal_2024}, smaller peaks are typically observed in the dynamics of $C(t)$, for squeezed coherent states~\cite{PhysRevA.40.6095, PhysRevA.36.1288, PhysRevA.47.4474, PhysRevA.47.4487, PhysRevA.34.3466, yi1997squeezed, EZAWA1991216, PhysRevA.40.2494, RevModPhys.58.1001, PhysRevA.47.5138, PhysRevA.61.010303, PhysRevA.61.022309, PhysRevLett.80.869, PhysRevA.60.937}, and the peaks around $\lambda t = 8$ are particularly prominent. However, in the presence of thermal photons, these smaller peaks vanish and the heights of the larger peaks decrease as well. This suggests that thermal photons contribute to the degradation of atom-atom entanglement in the system.

\subsection{Wigner function distributions for SCTS}
Before proceeding further to investigate the effects of different interactions, we discuss the Wigner function $W(\alpha)$ for SCTS in IDDJCM and DJCM. Wigner function distributions are of major importance in quantum optics. $W(\alpha)$ is defined as\cite{HILLERY1984121, schleich, agarwal2013}
\begin{equation}
W(\alpha)=\frac{1}{\pi^{2}}\int d^{2}\beta\enspace \text{Tr}[\hat{\rho}\hat{D}(\beta)]\exp(\beta^{*}\alpha-\beta\alpha^{*}).
\end{equation}
\begin{figure}
    \centering
    \includegraphics[scale = 0.4]{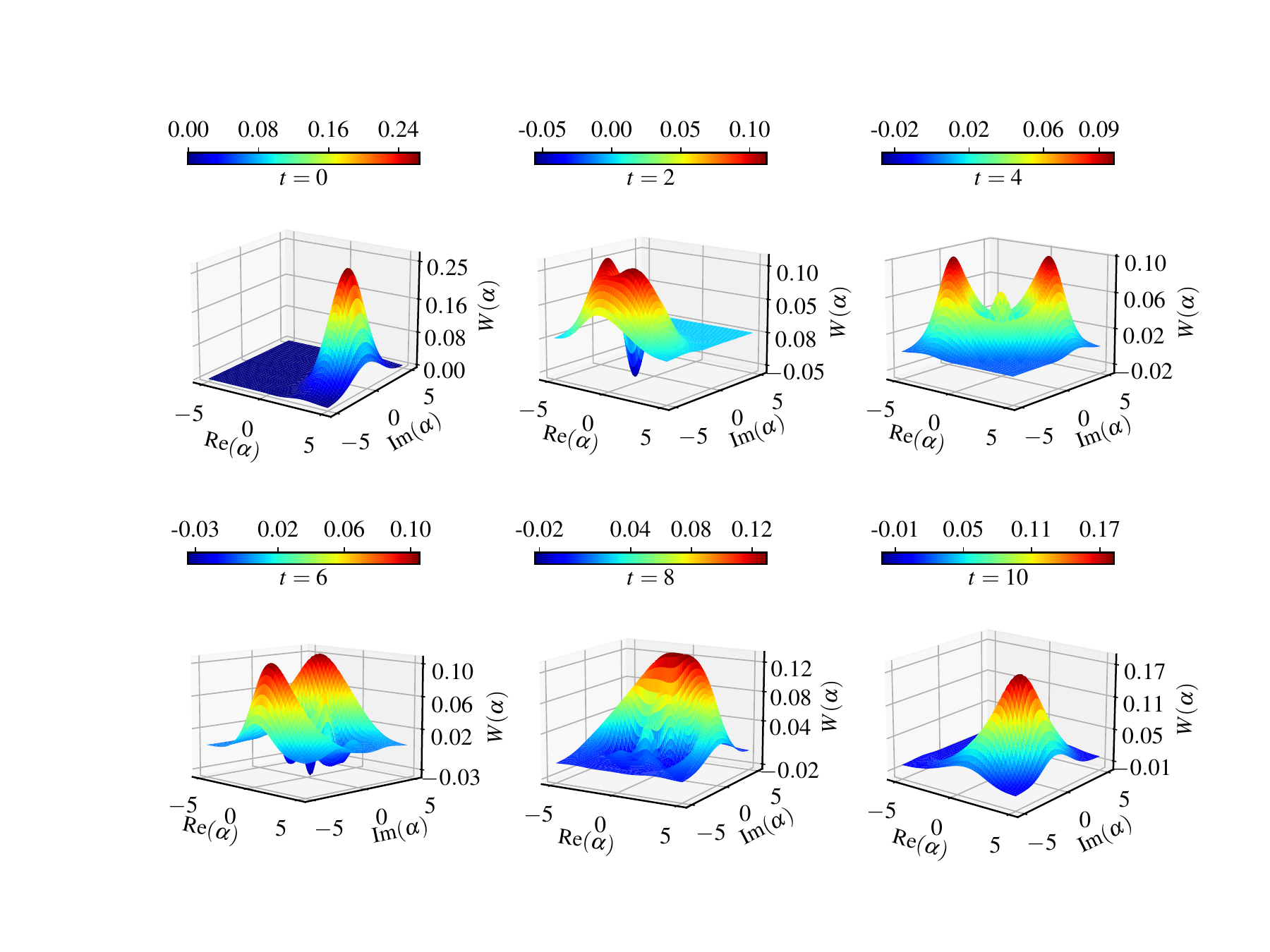}
    \caption{Wigner functions for SCTS in IDDJCM. The values of the parameters used in here are $\bar{n}_c = 2.0, \bar{n}_s = 0.1, \bar{n}_{th}=0.1$ and $\theta = \frac{\pi}{4}$.}
    \label{fig_3}
\end{figure}

\begin{figure}
    \centering
    \includegraphics[scale = 0.4]{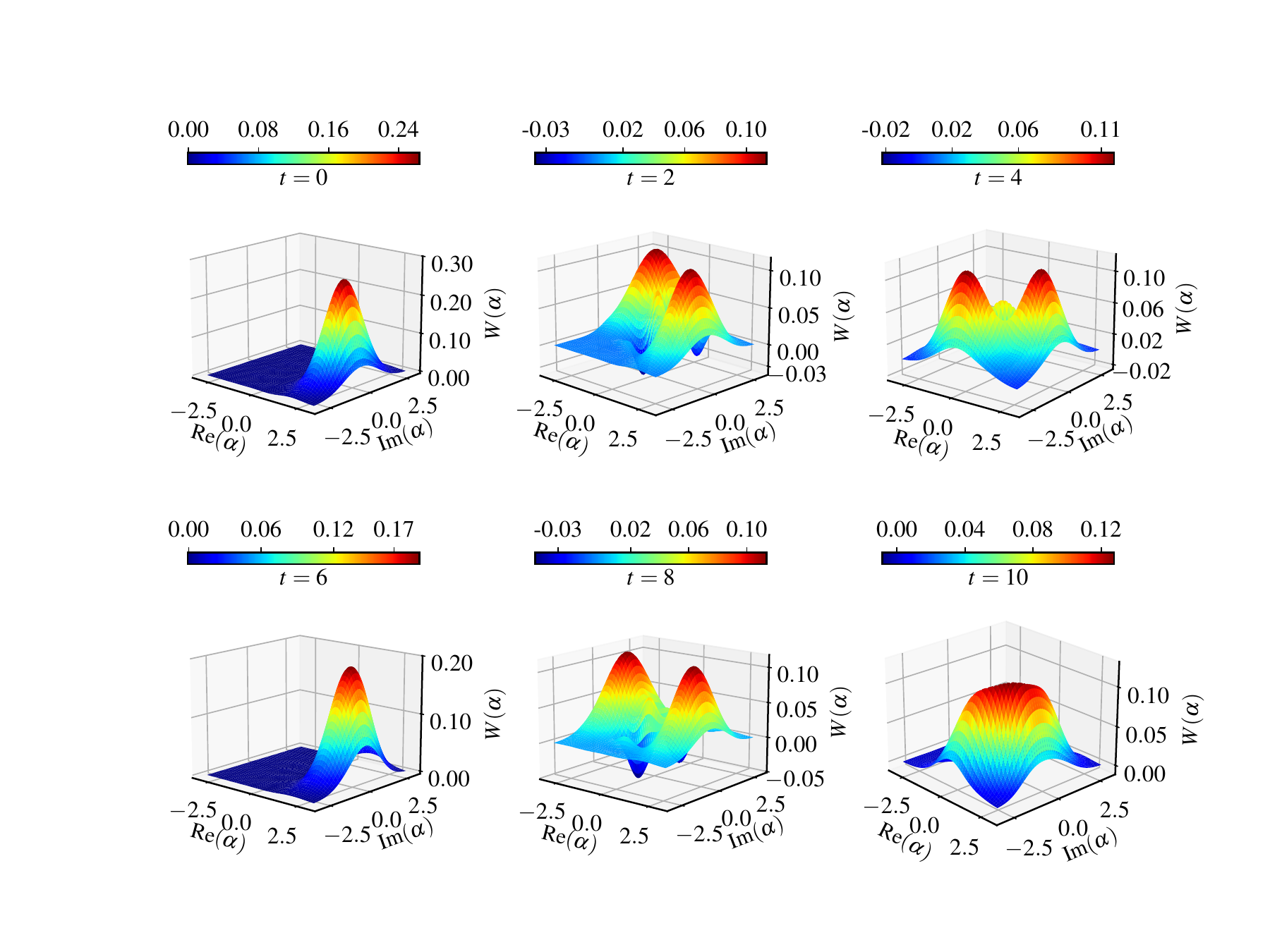}
    \caption{Wigner functions for SCTS in DJCM. The values of the parameters used in here are $\bar{n}_c = 2.0, \bar{n}_s = 0.1, \bar{n}_{th}=0.1$ and $\theta = \frac{\pi}{4}$.}
    \label{fig_4}
\end{figure}

The density operator $\hat{\rho}$ for the squeezed coherent thermal state (SCTS) is given in Eq.~(\ref{rho_scts}). The corresponding Wigner function $W(\alpha)$, which provides a quasi-probability distribution in phase space, is used as a reliable indicator of the quantum state's nonclassicality. In particular, the presence of negative regions in $W(\alpha)$ signals strong nonclassical behavior.

The Wigner function for IDDJCM and DJCM are presented in Figs. \ref{fig_3} and \ref{fig_4}. The contrasting behavior between the two models highlights an important physical insight: the intensity-dependent coupling in IDDJCM plays a stabilizing role, enabling temporary nonclassicality, and also permitting the system to re-enter a classical regime periodically. In DJCM, the absence of intensity dependence leads to more persistent dephasing and deeper excursions into nonclassical territory. Hence, while both models generate nonclassical light due to atom--field interactions, IDDJCM exhibits quasi-periodic modulation of classicality, whereas DJCM drives the system more deeply and irreversibly into nonclassical states.

These findings suggest that tuning interaction parameters and the nature of the coupling allows control over the quantum-to-classical transition and the degree of nonclassicality in the radiation field, which is crucial for quantum technologies relying on engineered quantum states of light.

\section{Effects of photon exchange interaction (PEI) on the entanglement dynamics}

Now we study the effects of single photon exchange between the two cavities on entanglement dynamics. Recent studies have explored scenarios where  photons can hop from one atom-cavity system to another. The effects of photon exchange interaction is modeled by the following Hamiltonian:
\begin{equation}
    \hat{H}_{\text{PE}} = \hat{H} + \kappa \hat{a}^{\dagger}{\hat{b}} + 
 \kappa \hat{b}^\dagger \hat{a},
\end{equation}
where $\kappa$ is the cavity-cavity photon exchange coupling term.
In particular, Pandit et. al., \cite{Pandit_2018, PhysRevA.101.053805} have studied the effects of cavity-cavity interaction on the entanglement dynamics of a generalized double Jaynes-Cummings model. It is shown that for larger value of $\kappa$, entanglement for the atom-atom subsystem can be protected. They have also investigated the role of $\kappa$ in entanglement transfer between atom-atom and field-field subsystems. In another work, Laha \cite{laha2023dynamics} has investigated the effects of beam-splitter which is equivalent to photon exchange interaction on entanglement dynamics in DJCM for radiation states in vacuum, coherent and thermal states. Without intensity dependence this Hamiltonian becomes
\begin{equation}
    \hat{H'}_{\text{PE}} = \hat{H'} + \kappa \hat{a}^{\dagger}{\hat{b}} + 
 \kappa \hat{b}^\dagger \hat{a}.
\end{equation}

\begin{figure}[ht]
    \centering
    \includegraphics[scale = 0.45]{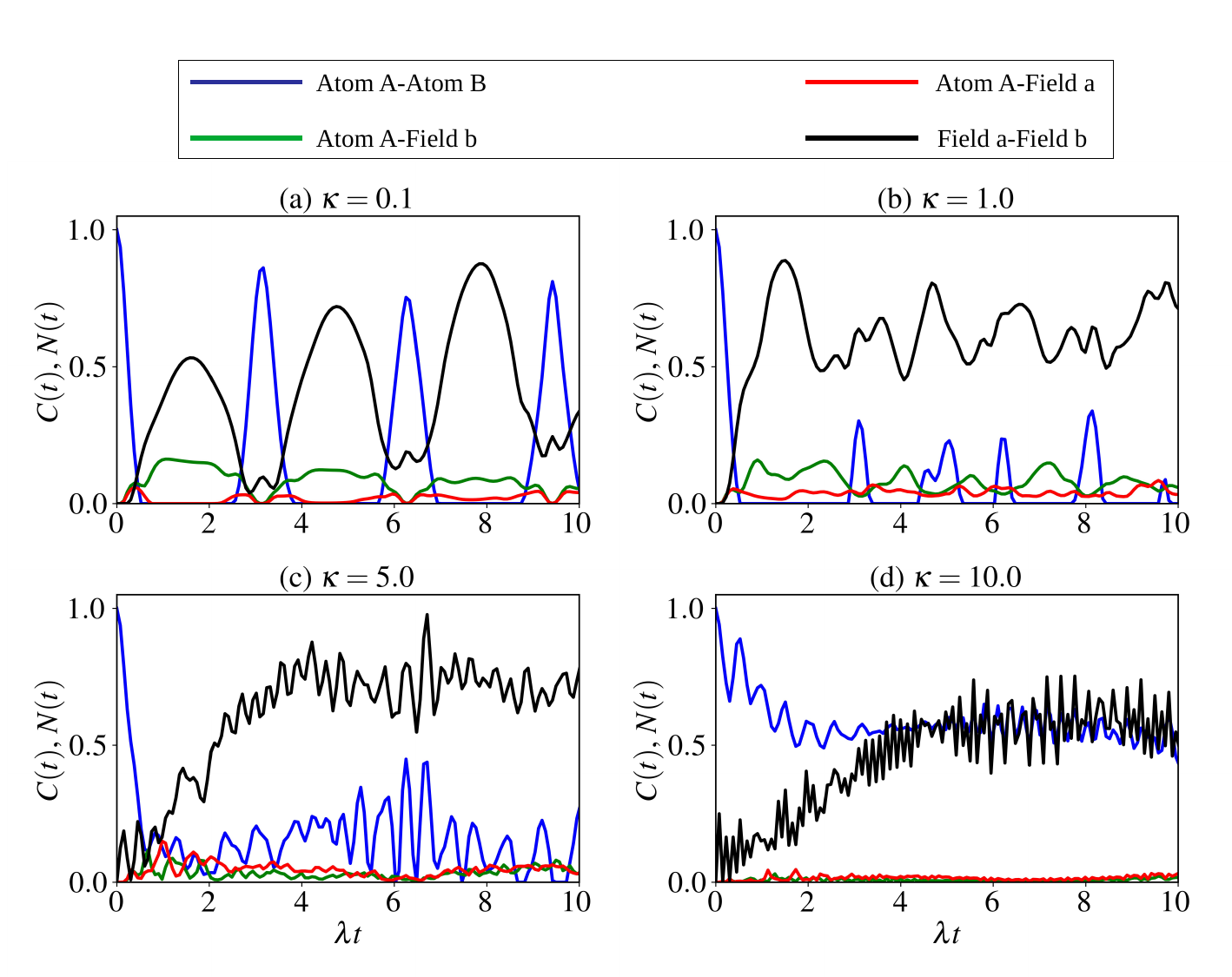}
    \caption{Effects of photon exchange interaction on entanglement dynamics for SCTS in IDDJCM. The values of the parameters used in these plots are $\bar{n}_c = 2$, $\bar{n}_{th} = 0.1$, $\bar{n}_s = 1.0$, $\kappa = 0.1, 1.0, 5.0, 10.0$ and $\theta = \frac{\pi}{4}$.}
    \label{fig_5}
\end{figure}

\begin{figure}[ht]
    \centering
    \includegraphics[scale = 0.45]{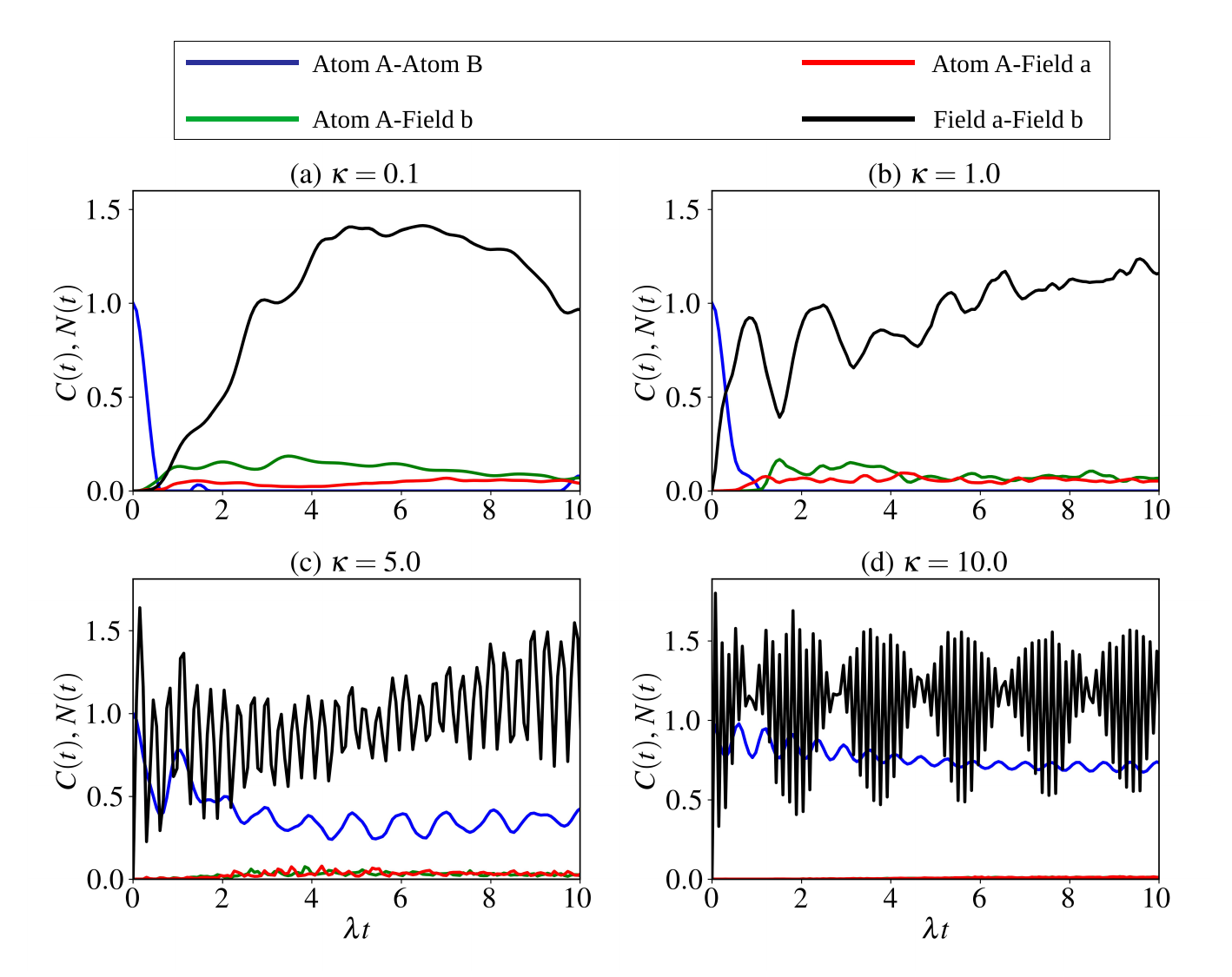}
    \caption{Effects of photon exchange interaction on entanglement dynamics for SCTS in DJCM. The values of the parameters used in these plots are $\bar{n}_c = 2$, $\bar{n}_{th} = 0.1$, $\bar{n}_s = 1.0$, $\kappa = 0.1, 1.0, 5.0, 10.0$ and $\theta = \frac{\pi}{4}$.}
    \label{fig_6}
\end{figure}

Figure \ref{fig_5} shows the effects of cavity-cavity single photon exchange interaction on the entanglement dynamics for SCTS in IDDJCM. The strength of this cavity-cavity interaction via single photon exchange is characterized by the parameter $\kappa$. To study the effects of $\kappa$, $\bar{n}_c = 2, \bar{n}_s = 1$ and $\bar{n}_{th} = 0.1$ are taken. From Fig. \ref{fig_6}(a), it is observed that for $\kappa = 0.1$, i.e., $\kappa < \lambda$, the peaks of atom A-atom B entanglement pulses decrease with time (described by blue curve); however, the lengths of ESDs do not change considerably. The field a-field b entanglement which is represented by the black curve shows different behaviour. If we compare Figs. \ref{fig_6}(d) and \ref{fig_1}(d), it can be seen that ESDs are removed from the dynamics and the amplitude of $N(t)$ increases with time. For both atom A-field a (green curve) and atom A-field b (red curve), the periodic nature of $N(t)$ gets destroyed; nevertheless, the amplitude of $N(t)$ does not change noticeably. The PEI allows photons to hop between cavities, effectively coupling the two field modes and disrupting their independent evolution. This interaction mixes the Fock components across the cavities, increasing photon number uncertainty and breaking the regular structure essential for coherent rephasing. As a result, the system loses its well-defined Rabi frequencies tied to fixed photon numbers, leading to spectral irregularity and enhanced dephasing. The entanglement dynamics become more complex and less periodic, with entanglement flowing not only within each atom–field pair but also across atom–atom and field–field subsystems, resulting in a more
irregular and intertwined evolution.

For $\kappa = \lambda$ (i.e., $\kappa = 1$), the peaks of the atom-atom entanglement $C(t)$ decay rapidly over time, and the duration of the first episode of entanglement sudden death (ESD) increases. However, the durations of subsequent ESD intervals tend to decrease in certain regions. In the case of field-field entanglement, the measure $N(t)$ exhibits a sharp initial rise followed by oscillatory behavior. For the hybrid subsystems—atom A with field a and atom A with field b—all instances of ESD are eliminated, although the amplitudes of $N(t)$ in these subsystems remain relatively low.

As the photon exchange coupling is increased further, with $\kappa > \lambda$ (e.g., $\kappa = 5$), ESD in the dynamics of $C(t)$ is removed, since entanglement appears to be nonvanishing (though it shows an irregular behaviour). These observations suggest that the photon exchange interaction actively generates atom-atom entanglement. Simultaneously, the amplitude of field-field entanglement $N(t)$ decreases, which can be attributed to the redistribution of entanglement from the field-field subsystem to the atom-atom subsystem. The frequency of oscillations in the field-field entanglement also increases significantly. Meanwhile, entanglement between atom A and field a diminishes and becomes nearly zero at certain times. The entanglement between atom A and field b does not change considerably, although it vanishes whenever atom-atom and field-field entanglements reach their maxima.

When the photon exchange strength is raised to $\kappa = 10$, placing the system in the regime $\kappa \gg \lambda$, the photon exchange interaction dominates over the atom-field coupling. In this scenario, all instances of ESD in the atom-atom entanglement dynamics are eliminated, and $C(t)$ increases significantly in magnitude. This enhancement in atom-atom entanglement corresponds to a marked decrease in field-field entanglement $N(t)$, indicating a transfer of entanglement from the field-field subsystem to the atom-atom subsystem. Furthermore, $N(t)$ for both atom A–field a and atom A–field b subsystems becomes negligible throughout most of the evolution, aside from some small initial peaks.

In summary, strong photon exchange coupling between cavities suppresses ESD in both atom-atom and field-field entanglement dynamics. While the amplitude of $C(t)$ increases notably with increasing $\kappa$, the amplitude of field-field entanglement $N(t)$ diminishes, illustrating a clear transfer of entanglement from the fields to the atoms as photon exchange becomes more dominant.

Figure~\ref{fig_7} illustrates the effects of photon exchange interaction on SCTS in the double Jaynes–Cummings model (DJCM). It is observed that the maximum value of $N(t)$ for the field-field subsystem increases by nearly three times due to the photon exchange interaction.
For $\kappa = 0.1$, the lengths of the entanglement sudden death (ESD) intervals in the atom-atom entanglement increase, and several smaller peaks in the dynamics disappear. The entanglement between atom A and field $a$ reduces significantly, and at certain times, $N(t)$ almost vanishes. In contrast, for the atom A–field $b$ subsystem, the ESDs are removed from the dynamics of $N(t)$. At  $\kappa =1$, the ESD duration in $C(t)$ becomes longer as the minor peaks in the dynamics are suppressed. The field-field entanglement $N(t)$ rises sharply and then decreases, although its maximum value becomes lower than that in the $\kappa = 0.1$ case. In the atom A–field $a$ subsystem, the increase in $\kappa$ induces ESD, but no significant change is observed in the atom A–field $b$ entanglement.

As $\kappa$ increases further to $5$, the entanglement dynamics of $C(t)$ undergo a drastic transformation. All ESDs are eliminated, and the value of $C(t)$ increases significantly at all times. For the field-field subsystem, the maximum value of $N(t)$ increases, and the entanglement dynamics become highly oscillatory. In both atom A–field $a$ and atom A–field $b$ subsystems, $N(t)$ exhibits initial ESD, but subsequently remains nonzero with very low amplitude. In the regime where $\kappa \gg 1$ (e.g., $\kappa = 10$), $C(t)$ increases further and maintains a high value throughout the evolution. Interestingly, the dynamics of $N(t)$ for the field-field entanglement reveal the formation of entanglement patterns, with significantly enhanced amplitudes. Meanwhile, the entanglement between atom A and both fields a and b vanishes completely.

From these observations, it can be concluded that photon exchange interaction primarily generates entanglement in the atom-atom and field-field subsystems. In the case of IDDJCM, we observe that $C(t)$ is on average higher while $N(t)$ for the field-field entanglement is on average lower with respect to DJCM. Conversely, in the standard DJCM, both $C(t)$ and field-field $N(t)$ increase with $\kappa$. In both models, however, the entanglement between atom A and fields a and b vanishes as the photon exchange coupling $\kappa$ increases.

\section{Effects of dipole-dipole interaction between the two atoms on entanglement}

\begin{figure}[ht!]
    \centering
    \includegraphics[scale = 0.45]{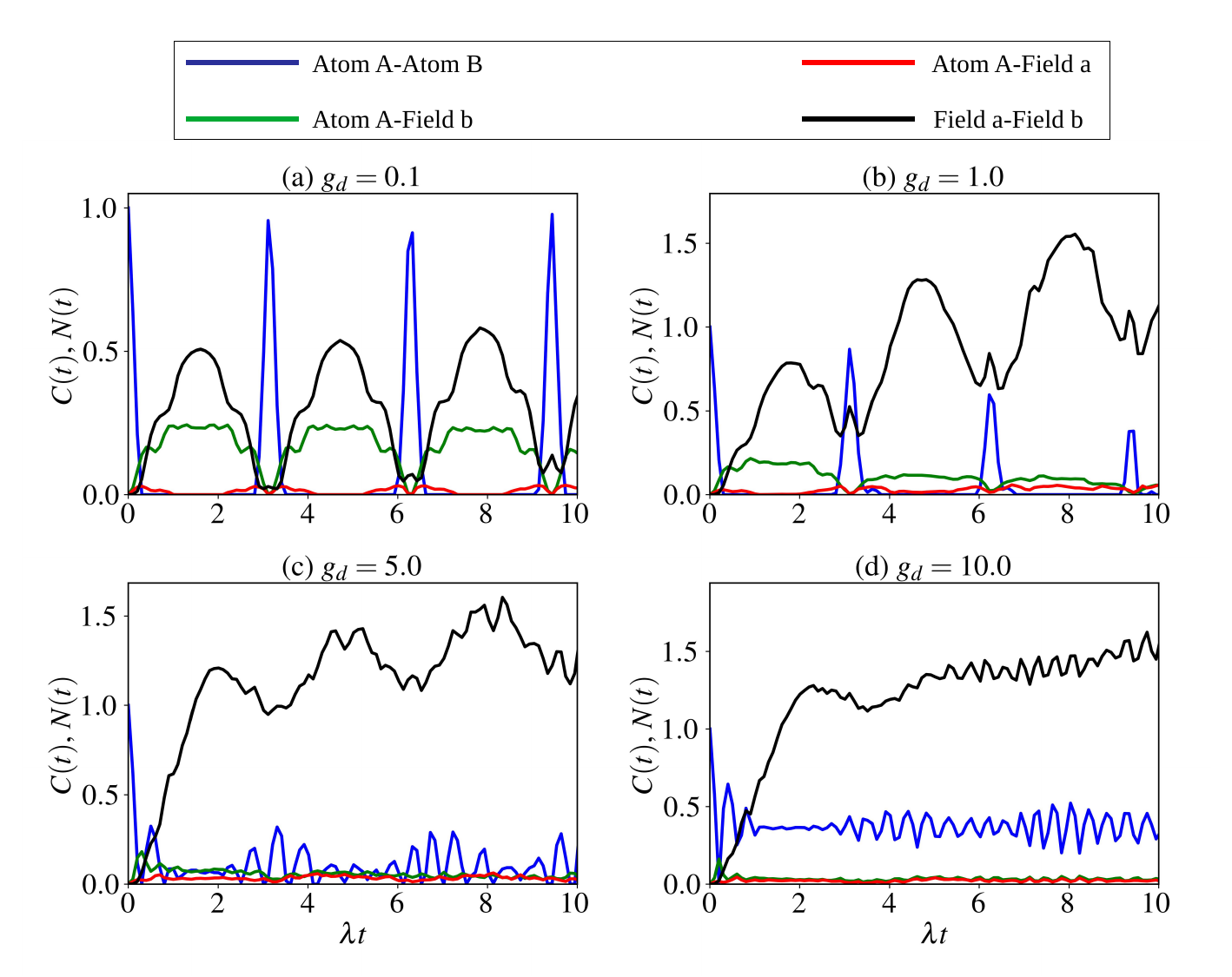}
    \caption{Effects of dipole-dipole interaction for SCTS in IDDJCM. The values of the parameters used in these plots are $\bar{n}_c = 2$, $\bar{n}_{th} = 0.1$, $\bar{n}_s = 1.0$, $g_d= 0.1, 1.0, 5.0, 10.0$ and $\theta = \frac{\pi}{4}$.}
    \label{fig_7}
\end{figure}
In this section, the effects of dipole-dipole interaction on the entanglement dynamics are discussed. The intensity-dependent double JC Hamiltonian with dipole-dipole interaction can be written as 

\begin{equation}
     \hat{H}_{\text{dd}} = \hat{H} + g_{d}(\hat{\sigma}^{\text{A}}_{+} \hat{\sigma}^{\text{B}}_{-} + \hat{\sigma}^{\text{A}}_{-} \hat{\sigma}^{\text{B}}_{+}),
\end{equation}
where $g_d$ is the dipole-dipole coupling strength. For the double Jaynes-Cummings model, this Hamiltonian becomes
\begin{equation}
     \hat{H'}_{\text{dd}} = \hat{H'} + g_{d}(\hat{\sigma}^{\text{A}}_{+} \hat{\sigma}^{\text{B}}_{-} + \hat{\sigma}^{\text{A}}_{-} \hat{\sigma}^{\text{B}}_{+}).
\end{equation}
Dipole-dipole interaction introduces a direct coupling between the atoms, independent of the field. It adds new energy terms (e.g., exchange of atomic excitation) that interfere with the field-mediated evolution. The resulting energy spectrum becomes anharmonic, leading to incommensurate Rabi frequencies and dephasing.

Much interest has gone into studying the effects of the dipole-dipole interaction on the dynamics of the atom-field system for a long time. In particular, the authors in \cite{PhysRevA.44.2135}, have discussed a detailed dynamics of the two-atom system where two atoms are interacting with each other via dipole-dipole coupling. A remarkable feature which comes out of this study is that more series of revivals appear in the dynamics that are like the one-atom case, partial and overlapping. It is also found that quantum collapse is no longer Gaussian and now depends on the dipole-dipole interaction parameter. In Ref.\cite{Evseev_2017}, Evseev et. al., have investigated the entanglement dynamics between two qubits in a non-resonant DJCM taking into account the direct dipole-dipole interaction between the qubits. The results show that the dipole-dipole parameter has great impact on the entanglement dynamics. They have also shown that the presence of large $g_d$ leads to a stabilization of entanglement for all Bell-type initial qubit states and different couplings and detunings. In another work \cite{Santos-Sánchez_2016} , the authors have studied the nonlinear version of the Jaynes-Cummings model for two identical two-level atoms for Ising-like and dipole-dipole interaction between the atoms. It is found that when the ratio  $\frac{g_d - J_z}{\lambda} >> 1$, certain significant results are observed. The effects of dipole-dipole interaction on the energy levels of an one-dimensional system using Jaynes-Cummings Hubbard model is discussed in Ref.\cite{PhysRevE.106.064107}.

\begin{figure}[ht]
    \centering
    \includegraphics[scale = 0.45]{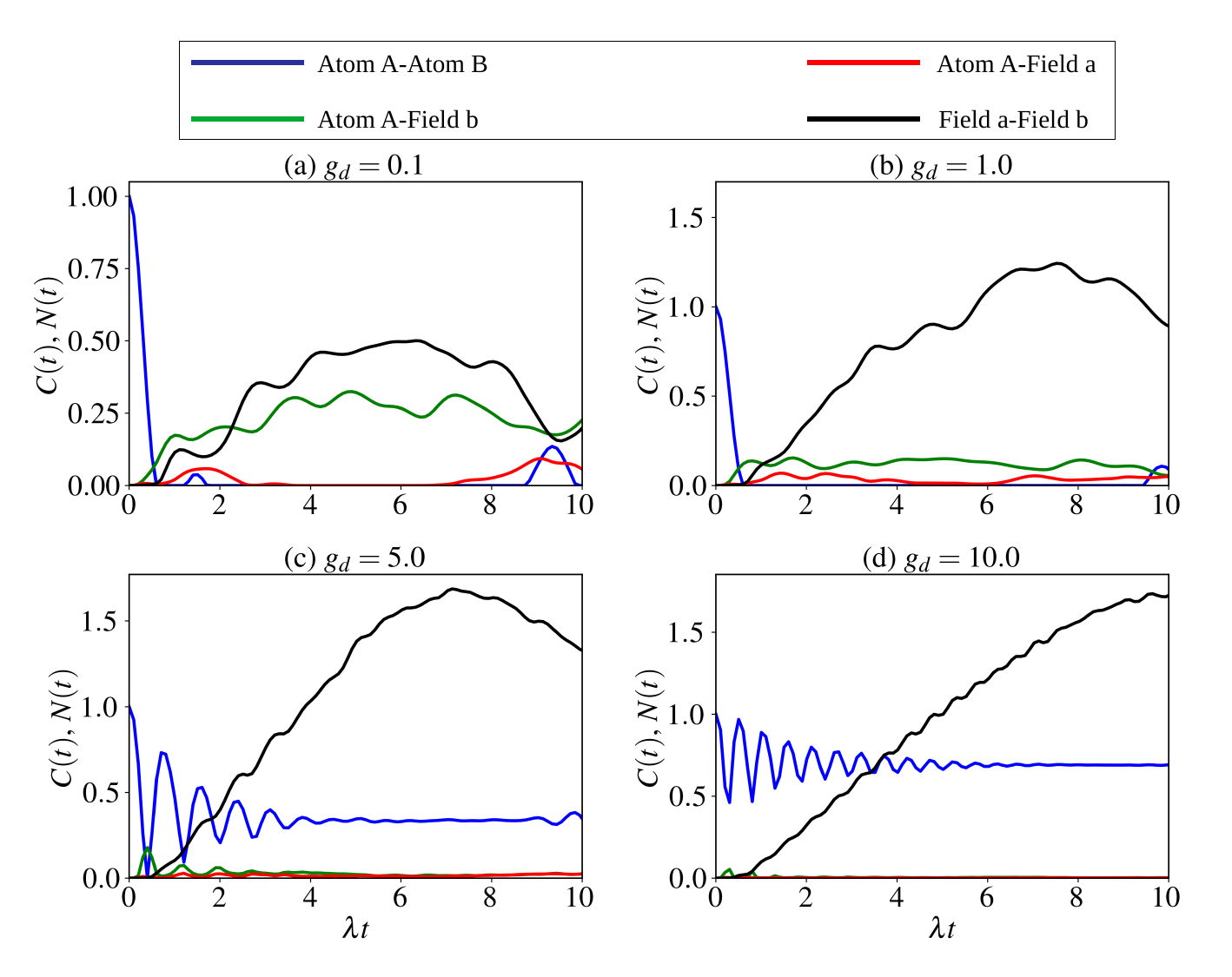}
    \caption{Effects of dipole-dipole interaction on entanglement dynamics for SCTS in DJCM. The values of the parameters used in these plots are $\bar{n}_c = 2$, $\bar{n}_{th} = 0.1$, $\bar{n}_s = 1.0$, $g_d= 0.1, 1.0, 5.0, 10.0$ and $\theta = \frac{\pi}{4}$.}
    \label{fig_8}
\end{figure}
The effects of dipole--dipole interaction (DDI) between the atoms on the entanglement dynamics for the intensity-dependent double Jaynes--Cummings model (IDDJCM) are presented in Fig.~\ref{fig_7}. The DDI strength is denoted by $g_d$. For lower value such as $g_d = 0.1$, the entanglement dynamics remain nearly unchanged. This is expected, as weak DDI has minimal influence on the coherent structure of the atom-field subsystems.

However, as the interaction strength increases to $g_d = 5$, notable changes emerge. The field--field entanglement becomes significantly enhanced in amplitude, though the characteristic oscillations disappear, indicating a shift toward more static field correlations. This reflects the fact that strong DDI enables more effective entanglement transfer between nonlocal subsystems, reducing oscillatory exchange with the local atom--field modes. At $g_d = 10$, the impact is even more pronounced. All entanglement sudden deaths (ESDs) are removed from the atom--atom concurrence $C(t)$, and while oscillations still occur initially, they gradually damp out over time. The field--field entanglement dynamics stabilize, halting the downward decay typically seen at weaker couplings. In contrast, the atom--field entanglements decay and nearly vanish. This behavior can be interpreted as a consequence of monogamy of entanglement: the strong nonlocal atomic interaction dominates the system dynamics and draws entanglement away from the local atom--field pairs.

The effects of DDI on the standard DJCM with SCTS are shown in Fig.~\ref{fig_8}. For $g_d = 0.1$, all subsystems show dynamics similar to the uncoupled case, except the field--field entanglement, which rises in amplitude over time and eliminates ESDs. At $g_d = 1.0$, the peaks of $C(t)$ are significantly reduced, although the ESD durations remain unchanged. Meanwhile, the field--field entanglement increases further and fully eliminates its own ESDs, suggesting a robust buildup of nonlocal photonic correlations. The amplitude of the atom A--field a entanglement $N(t)$ decreases, whereas $N(t)$ for atom A--field b remains largely unaffected.

For $g_d = 5$, new peaks emerge in $C(t)$, indicating the onset of additional coherent excitation pathways enabled by the DDI. At $g_d = 10$, all ESDs in the atom--atom subsystem are fully removed, and the field--field entanglement remains persistently high. However, the atom--field entanglements for both atoms are almost entirely suppressed.

These results highlight the physical role of the dipole--dipole interaction as a mechanism that redistributes entanglement from local (atom--field) to nonlocal (atom--atom and field--field) subsystems. As the DDI strength increases, the system transitions from local coherence-dominated dynamics to a regime where nonlocal correlations dominate, demonstrating a clear manifestation of entanglement monogamy and redistribution.

\section{Effects of spin-spin Ising interaction between the two atoms on entanglement}

In the present section, we study the effects of Ising-type interaction between the two atoms on the entanglement dynamics. The Hamiltonian with Ising-type interaction for IDDJCM and DJCM can be written as follows:

\begin{equation}
    \hat{H}_{\text{IS}} = \hat{H} + J_{z}\, \hat{\sigma}^{\text{A}}_{z} \otimes \hat{\sigma}^{\text{B}}_{z},
\end{equation}
and
\begin{equation}
    \hat{H'}_{\text{IS}} = \hat{H'} + J_{z}\, \hat{\sigma}^{\text{A}}_{z} \otimes \hat{\sigma}^{\text{B}}_{z},
\end{equation}
where $J_z$ is the coupling strength between the two atoms. $J_z$ has the unit of energy. Recently,there are studies on the Ising-type interaction for different configurations. Ghoshal et. al., have investigated the entanglement dynamics of the quenched disordered double Jayens-Cummings model in the presence of Ising-type interaction in the Hamiltonian\cite{PhysRevA.101.053805}. In another work, Pandit et. al., have also studied the effects of Ising-type interaction on the entanglement dynamics in the DJCM \cite{Pandit_2018}.  In Ref.\cite{laha2023dynamics}, Laha has analyzed the effects of Ising-type interaction on entanglement in DJCM for vacuum, coherent and thermal states of radiation fields. Sadiek et. al.,\cite{e23050629} have investigated the time evolution and asymptotic behaviour of entanglement with including the dipole-dipole and Ising-type interactions in the Hamiltonian.

\begin{figure}[ht]
    \centering
    \includegraphics[scale = 0.45]{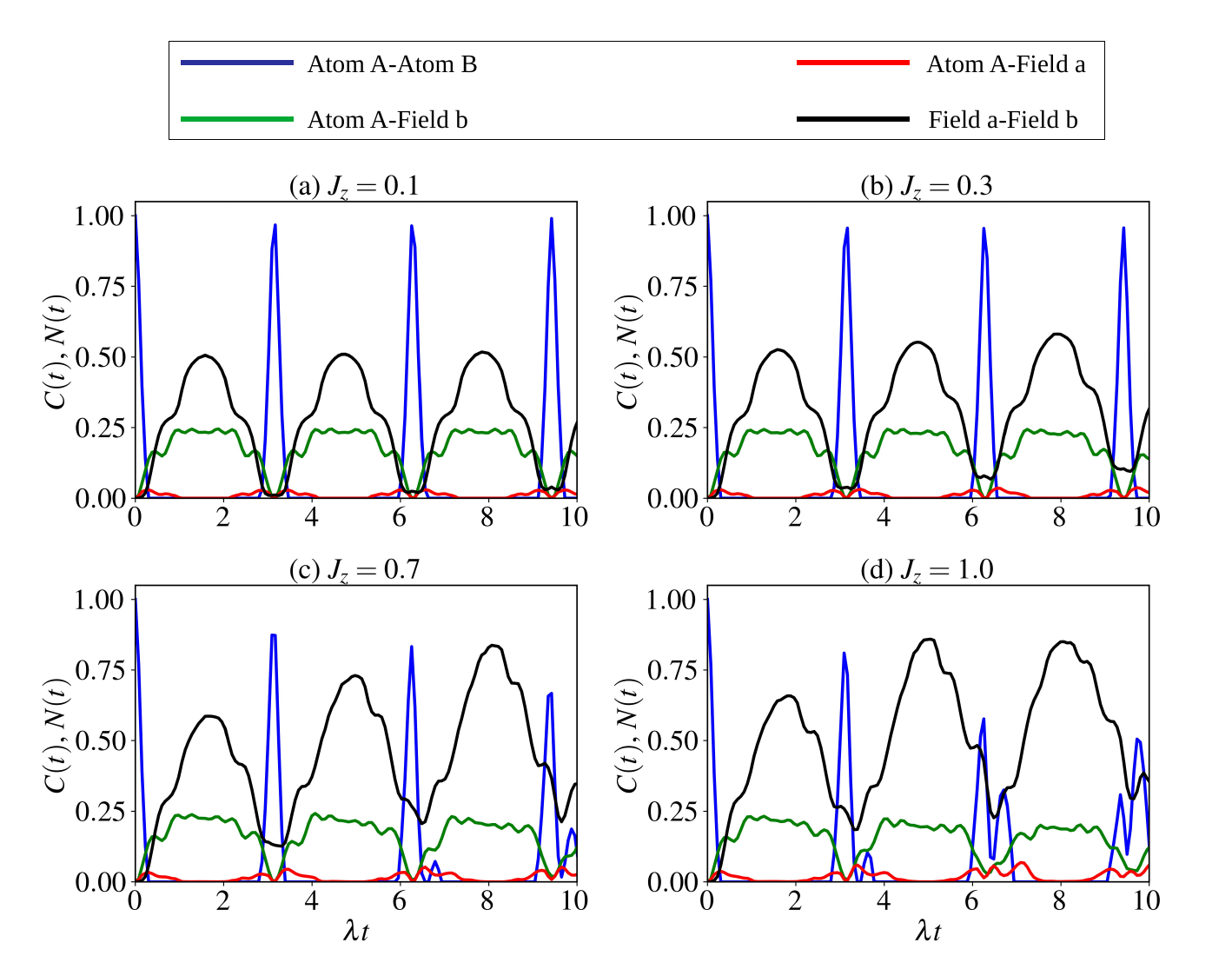}
    \caption{Effects of Ising interaction on entanglement dynamics in IDDJCM. The values of the parameters used in these plots are $\bar{n}_c = 2$, $\bar{n}_{th} = 0.1$, $\bar{n}_s = 1.0$, $J_{z} = 0.1, 0.3, 0.7, 1.0$ and $\theta = \frac{\pi}{4}$.}
    \label{fig_9}
\end{figure}

The influence of Ising-type interaction on the entanglement dynamics for the intensity-dependent double Jaynes--Cummings model (IDDJCM) is illustrated in Fig.~\ref{fig_9}. Notably, the behavior observed here differs significantly from the corresponding dynamics in the standard DJCM for squeezed coherent states (SCS) and Glauber--Lachs (G--L) states\cite{Mandal_2024}. As shown in Figs.~\ref{fig_9}(a) and \ref{fig_9}(b), for small values of the Ising interaction strength, namely \( J_z = 0.1 \) and \( 0.3 \), the entanglement dynamics of the atom--atom subsystem, \( C(t) \), and the atom--field entanglement, \( N(t) \), remain nearly indistinguishable from the case with \( J_z = 0 \) (see Figs.~\ref{fig_1} and \ref{fig_9}). However, subtle effects are observed in the cross-coupled subsystems. Specifically, the entanglement between atom A and field \( b \) shows a reduction in the amplitude of \( N(t) \), along with the persistence of entanglement sudden deaths (ESDs). In contrast, the field \( a \)--field \( b \) entanglement shows notable enhancement: the amplitude increases and ESDs are eliminated, suggesting that the Ising interaction facilitates entanglement redistribution into the field--field sector.

As the Ising interaction strength increases to \( J_z = 0.7 \), new features emerge in the atom--atom entanglement dynamics. Small secondary peaks begin to appear in \( C(t) \), accompanied by a reduction in the overall amplitude. For the atom A--field \( a \) subsystem, \( N(t) \) shows a mild decrease in amplitude, while ESDs are gradually removed. Interestingly, the atom A--field \( b \) dynamics remain largely unaffected. However, the field--field entanglement continues to grow in amplitude and shows no signs of ESD, indicating stronger nonlocal correlations mediated by the Ising interaction. At \( J_z = 1.0 \), the small peaks in \( C(t) \) appear earlier in the evolution and grow over time, whereas the initial large peak diminishes, resulting in a net reduction in the duration of ESDs. In atom--field entanglements, the peak heights decrease for atom A--field \( a \), but increase for atom A--field \( b \) and field \( a \)--field \( b \) subsystems.

\begin{figure}[ht]
    \centering
    \includegraphics[scale = 0.45]{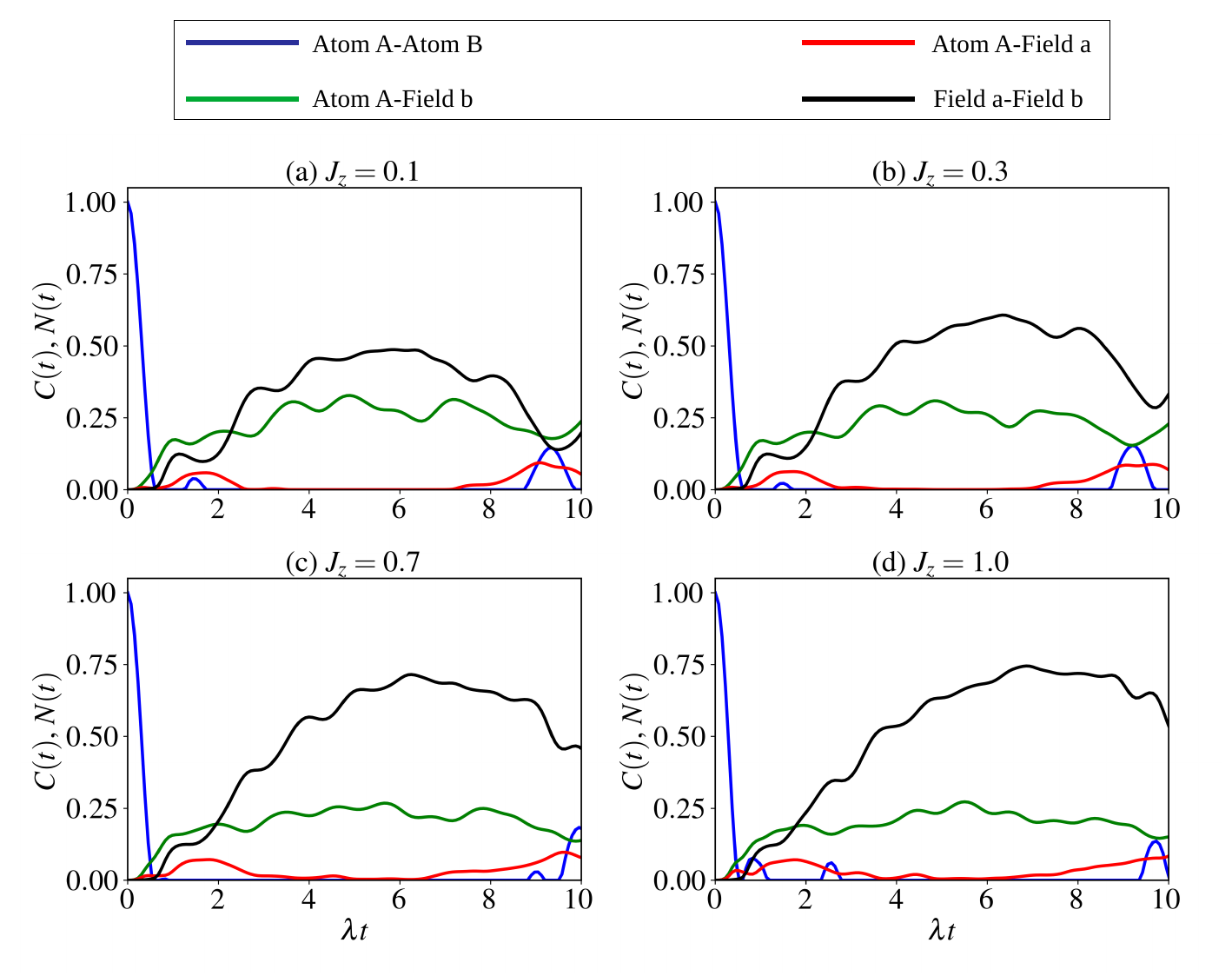}
    \caption{Effects of Ising interaction on entanglement dynamics in DJCM. The values of the parameters used in these plots are $\bar{n}_c = 2$, $\bar{n}_{th} = 0.1$, $\bar{n}_s = 1.0$, $J_{z} = 0.1, 0.3, 0.7, 1.0$ and $\theta = \frac{\pi}{4}$. }
    \label{fig_10}
\end{figure}
The effects of Ising interaction on entanglements in DCJM are depicted in Fig. \ref{fig_10}. In this case also, we observe that for small values of $J_z$ (0.1 and 0.3), $C(t)$ and $N(t)$ for all subsystems do not change considerably. Only field-field entanglement increases slightly for $J_z = 0.3$. For $J_z = 0.7$, length of ESD for $C(t)$ increases slightly, as the small bump disappears from the dynamics. The amplitudes of atom-field entanglement do not change noticeably; however, the length of ESD decreases for atom A-field b entanglement. For $J_z = 1.0$, two small peaks appear in the dynamics of $C(t)$ which makes the length of ESDs smaller. All other entanglement dynamics remain almost unchanged. 

These observations suggest that the Ising-type interaction introduces subtle redistributions of entanglement among the subsystems, particularly enhancing nonlocal correlations in the field sector. However, it is not significantly effective in eliminating ESDs in the atom--atom or cross atom--field subsystems. The nonlinearity introduced by the Ising term, while capable of modifying local coherence and interference effects, does not dominate the entanglement evolution unless its strength is significantly increased.

\section{Effects of Kerr-nonlinearity on the entanglement dynamics}

In this section, the effects of Kerr-nonlinearity on entanglement dynamics for IDDJCM and DJCM are investigated. The Hamiltonian of atom-field system with Kerr-nonlinearity is 

\begin{equation}
    \hat{H}_{\text{Kerr}} = \hat{H} + \chi \hat{a}^{\dagger 2} {\hat{a}}^{2} + 
\chi \hat{b}^{\dagger 2} \hat{b}^{2},
\end{equation}

\begin{equation}
    \hat{H'}_{\text{Kerr}} = \hat{H'} + \chi \hat{a}^{\dagger 2} {\hat{a}}^{2} + 
\chi \hat{b}^{\dagger 2} \hat{b}^{2},
\end{equation}
where $\chi = k \omega$ is the nonlinear coupling constant and $k$ is a non-negative number. Studying the effects of nonlinearity in the system has drawn attention in the past and in recent years \cite{PhysRevA.45.6816, PhysRevA.45.5056, PhysRevA.44.4623, ahmed2009dynamics, sivakumar2004nonlinear, Mo_2022, PhysRevB.105.245310, baghshahi2014entanglement, zheng2017intrinsic, PhysRevA.93.023844}. 

Kerr-nonlinearity affects the dynamics of atom-field interaction and other quantum optical quantities such as Q-funtion, Wigner function \cite{PhysRevA.44.4623}etc. significantly. In Ref.\cite{mojaveri2018thermal}, the authors have studied the effects of Kerr-nonlinearity and atom-atom coupling on the degree of atom-atom entanglement. In another work, Thabet et. al., \cite{thabet2019dynamics} have investigated the effects of Kerr-nonlinearity on the mean photon number, Mandel's $Q$ parameter, entropy squeezing and entanglement dynamics using nonlinear Jaynes-Cummings model. Xi-Wen Hou et. al., \cite{HOU2006727} have studied the dynamical properties of quantum entanglement in the integrable Jaynes-Cummings model with Kerr-nonlinearity with various Kerr coupling parameters and initial states, where the initial states are prepared by the coherent states placed in the corresponding phase space described in terms of canonical variables.

\begin{figure}[ht!]
    \centering
    \includegraphics[scale = 0.32]{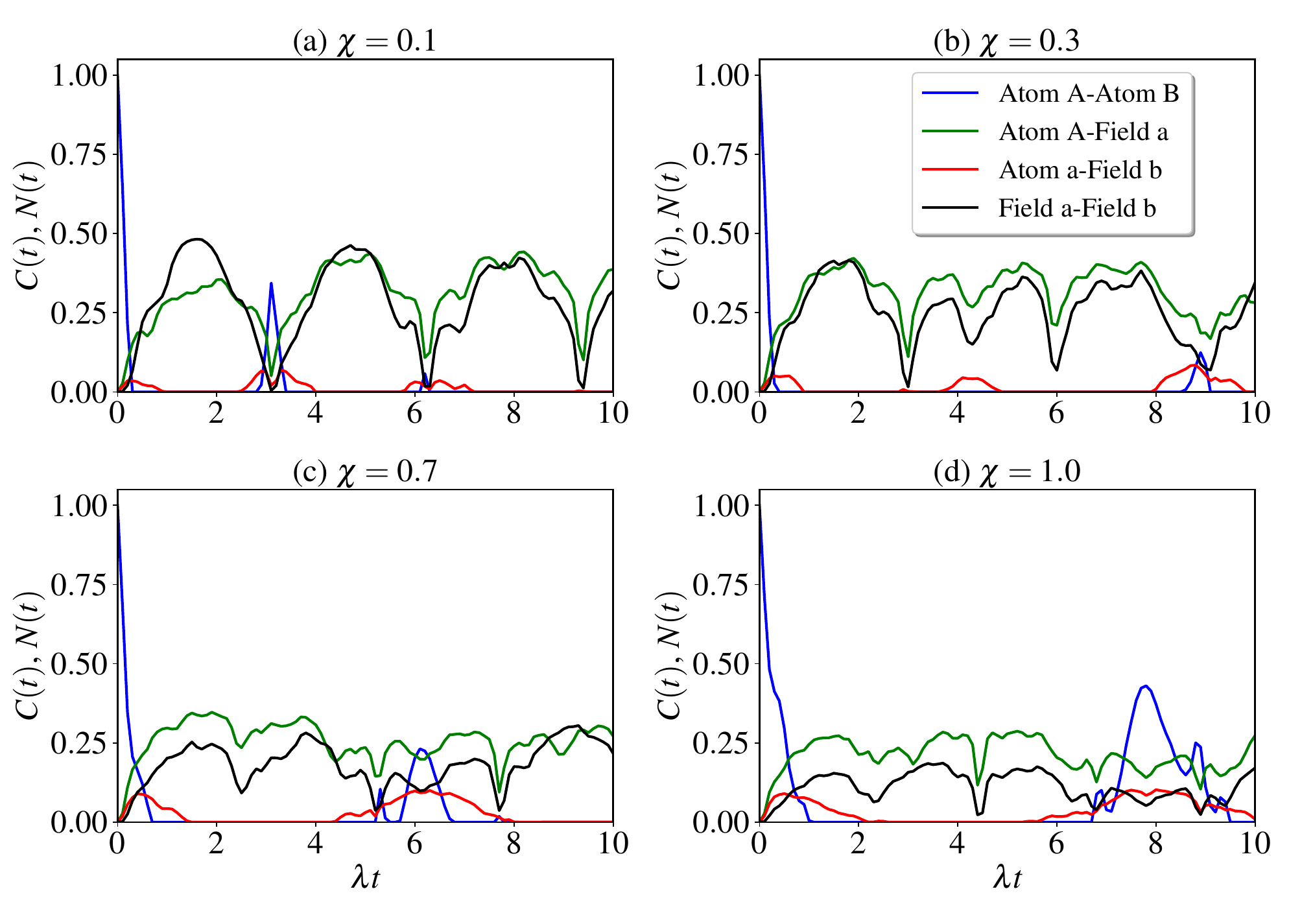}
    \caption{Effects of Kerr-nonlinearity on the entanglement dynamics for SCTS in IDDJCM with atoms in a Bell state. The values of the parameters used in these plots are $\bar{n}_c = 2$, $\bar{n}_{th} = 0.1$, $\bar{n}_s = 1.0$, $\chi = 0.1, 0.3, 0.7, 1.0$ and $\theta = \frac{\pi}{4}$.}
    \label{fig_11}
\end{figure}

The entanglement dynamics after adding Kerr-nonlinearity for IDDJCM are presented in Fig. \ref{fig_11}. The addition of nonlinearity $\chi = 0.1$ totally destroys the periodic nature of the entanglement for all the subsystem (blue curve in the plots). ESDs from the dynamics are removed for atom A-field a, atom A-atom b and field a-field b subsystems; however, in the case of atom A-atom B, ESDs become longer and peak heights get reduced significantly. Addition of Kerr-nonlinearity in the system modifies the energy levels of the field depending on photon number, introducing non-uniform spacing between adjacent levels. As $\chi$ is increased to $0.3$, ESDs get longer in the dynamics of $C(t)$ and also peak which were present till then disappear. For other subsystems, amplitude of $N(t)$ for atom A-field a (green curve) and field-field (black curve) entanglement decrease;  however, for atom A-field b subsystem, $N(t)$ increases. This happens because, the field b retain is retaining more structure or coherence under Kerr distortion, allowing it to maintain or enhance entanglement with atom A. For $\chi = 0.7$, which is represented by the red curve in the plots, we observe that $C(t)$ begins to revive again, making the lengths of ESDs shorter. In atom A-field a and field a-field b subsystems, amplitude $N(t)$ decreases significantly, though there is no ESD present in the system. Like atom-atom entanglement, $N(t)$ increases for atom A-field b subsystem and length of ESDs decreases. Though, $N(t)$ decreases at the beginning of the dynamics, it amplitude increases with time. So, it can be concluded that there is a critical value $\chi$ for which $C(t)$ becomes minimum and beyond that value it increases again (see \ref{app_E} for detail). As we increase the nonlinearity further to $\chi = 1.0$, same pattern (black curve) is observed as we see for $\chi = 0.7$ i.e., the atom-atom and atom A-field b entanglement increase while field-field and atom A-field a entanglement decrease further. The graphs indicate that the presence of Kerr nonlinearity favors the preservation of atom–atom and atom A–field b entanglement. This behavior suggests that Kerr nonlinearity redistributes quantum correlations among the subsystems, resulting in a competition for entanglement. Such redistribution is consistent with the principle of monogamy of entanglement, where the enhancement of entanglement in one pair limits its presence in others.

\begin{figure}[ht!]
    \centering
    \includegraphics[scale = 0.32]{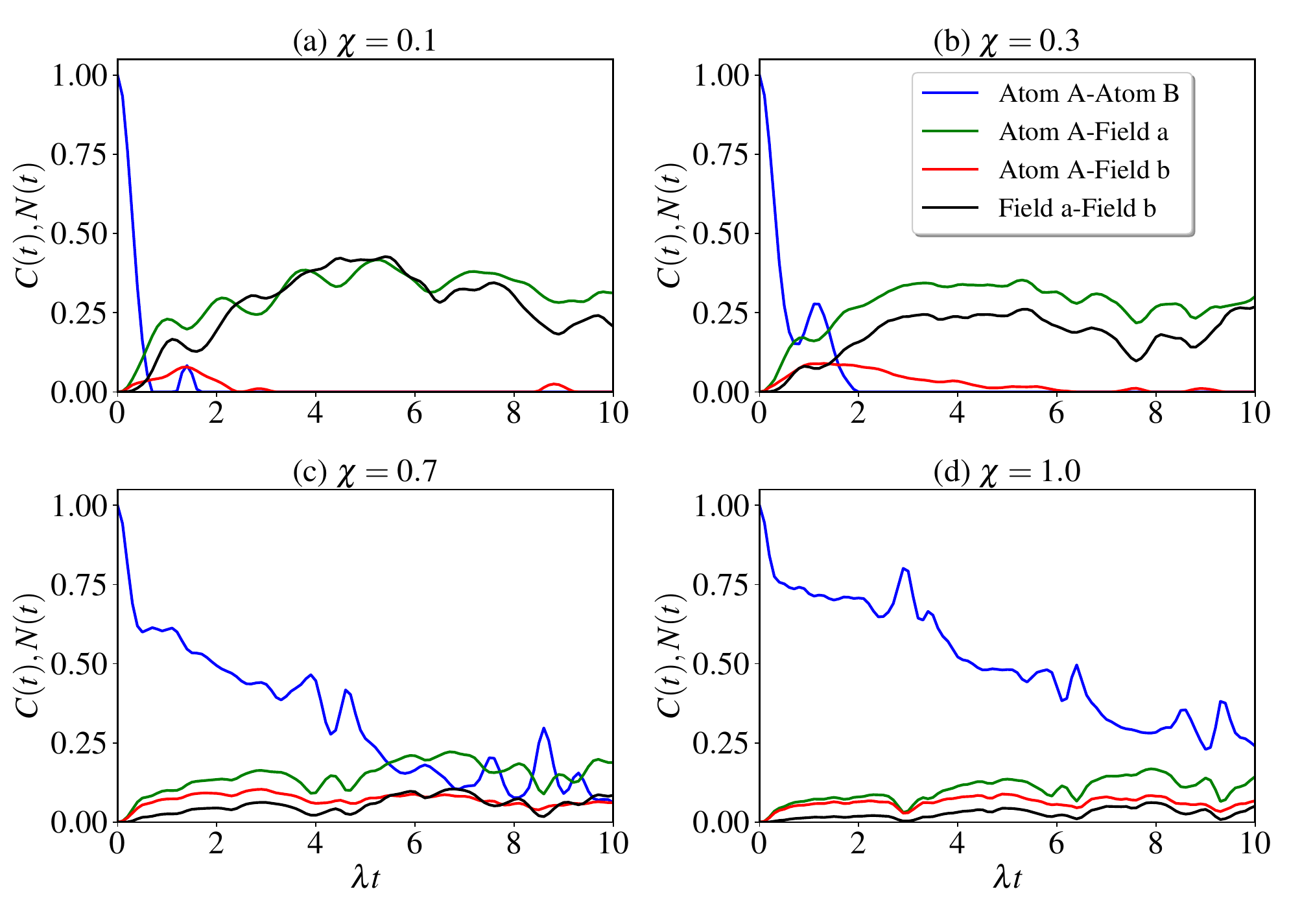}
    \caption{Effects of Kerr-nonlinearity on the entanglement dynamics for SCTS in DJCM with atoms in a Bell state.The values of the parameters used in these plots are $\bar{n}_c = 2$, $\bar{n}_{th} = 0.1$, $\bar{n}_s = 1.0$, $\chi = 0.1, 0.3, 0.7, 1.0$ and $\theta = \frac{\pi}{4}$.}
    \label{fig_12}
\end{figure}

The effects of Kerr-nonlinearity \((\chi)\) on the entanglement dynamics of various subsystems in the DJCM with squeezed coherent thermal states (SCTS) are depicted in Fig.~\ref{fig_12}. For \(\chi = 0.1\), the concurrence \(C(t)\) for the atom–atom subsystem exhibits a significant increase in the duration of entanglement sudden death, as the smaller revival peaks disappear (Fig.~\ref{fig_12}(a), blue curve). In contrast, the entanglement dynamics of the other subsystems, as measured by the negativity \(N(t)\), remain largely unaffected at this low Kerr strength. As \(\chi\) increases to 0.3, the amplitudes of both \(C(t)\) and \(N(t)\) rise slightly; however, ESDs persist in the dynamics. Notably, atom A–field a and field a–field b entanglements decrease in amplitude, while atom A–field b entanglement increases, with ESDs being completely removed from its dynamics. This suggests the onset of entanglement redistribution driven by Kerr-induced spectral distortion.

Upon further increasing \(\chi\) to 0.7 and 1.0, the concurrence \(C(t)\) exhibits a substantial enhancement in amplitude, with complete removal of ESDs. A similar trend is observed for the atom A–field b entanglement, indicating a strong transfer of quantum correlations into these subsystems. Conversely, entanglement in the atom A–field a and field–field subsystems is significantly diminished.

These results indicate that in this case also, the Kerr-nonlinearity acts as a mechanism for entanglement redistribution within the system, favoring the atom–atom and atom A–field b subsystems at the expense of others. Moreover, the enhancement of \(C(t)\) in the DJCM is observed to be more pronounced compared to the IDDJCM case.

\section{Effects of detuning on the entanglement}

\begin{figure}[ht]
    \centering
    \includegraphics[scale = 0.35]{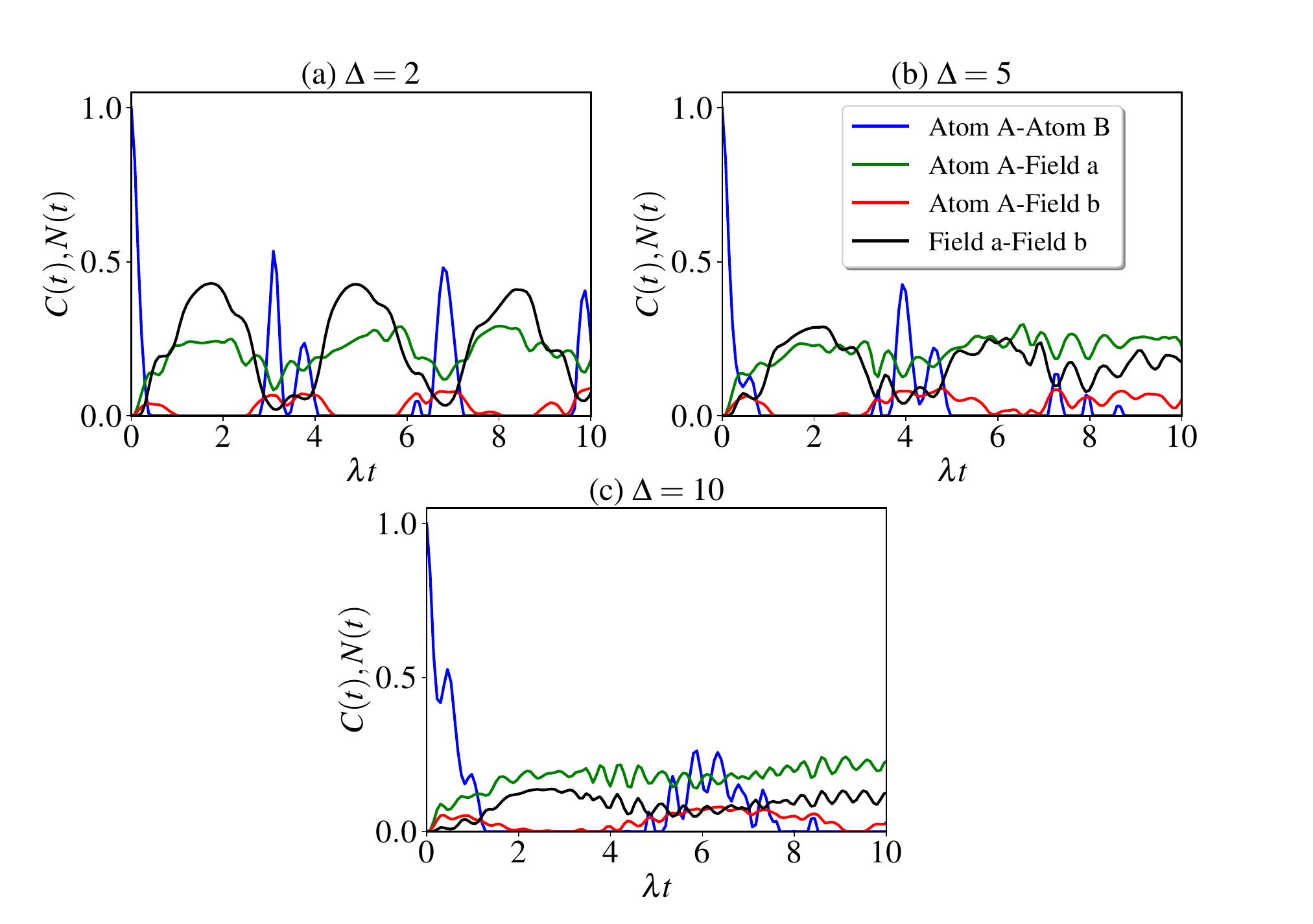}
    \caption{Effects of detuning on entanglement dynamics for SCTS in IDDJCM. The values of the parameters used in these plots are $\bar{n}_c = 2$, $\bar{n}_{th} = 0.1$, $\bar{n}_s = 1.0$, $\Delta = 2, 5, 10$ and $\theta = \frac{\pi}{4}$. }
    \label{fig_13}
\end{figure}

\begin{figure}[ht]
    \centering
    \includegraphics[scale = 0.35]{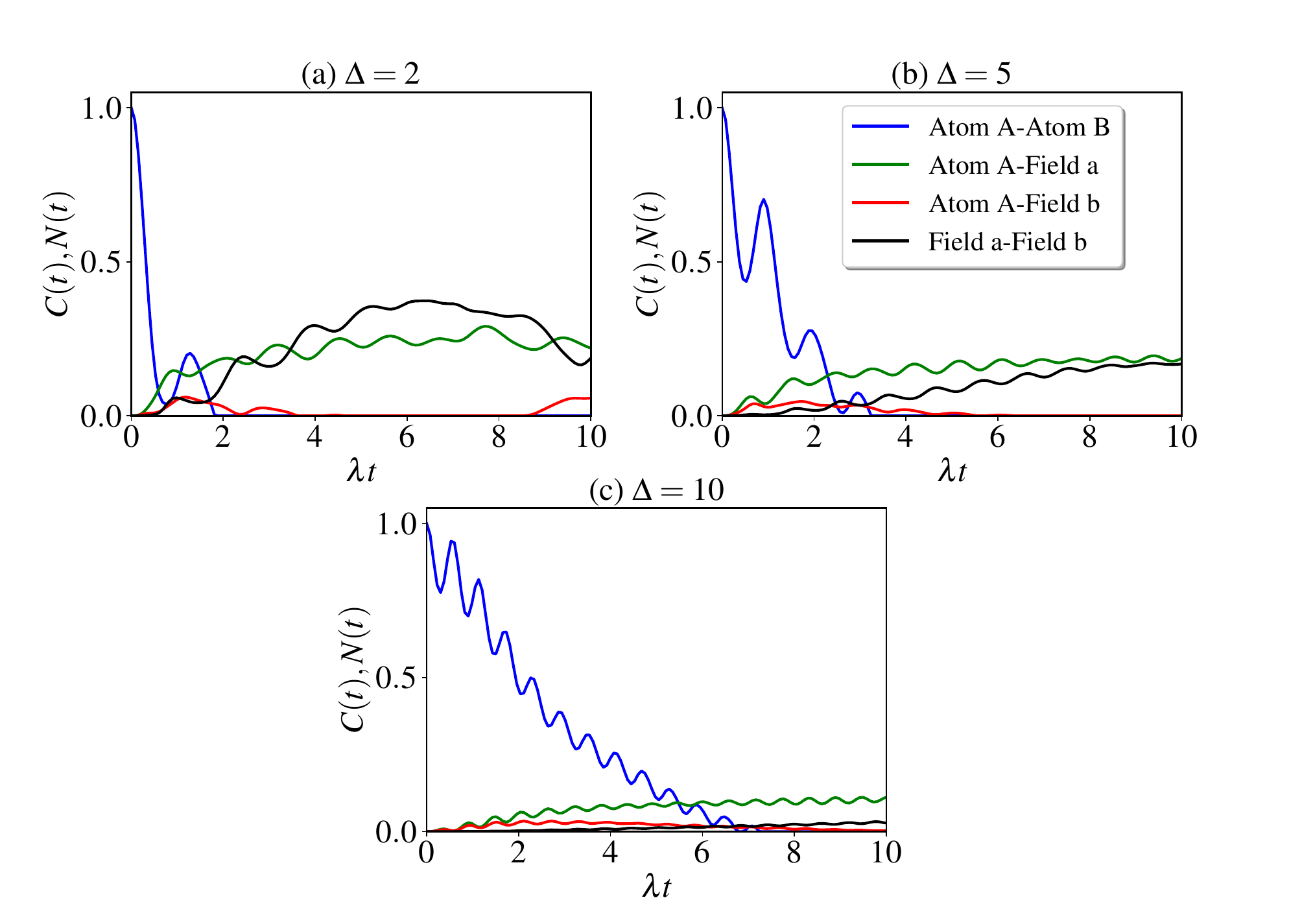}
    \caption{Effects of detuning on entanglement dynamics for SCTS in DJCM. The values of the parameters used in these plots are $\bar{n}_c = 2$, $\bar{n}_{th} = 0.1$, $\bar{n}_s = 1.0$, $\Delta = 2, 5, 10$ and $\theta = \frac{\pi}{4}$.}
    \label{fig_14}
\end{figure}

Now, we discuss the effects of detuning on the entanglement dynamics of various subsystems. Detuning is defined as $\Delta = \nu - \omega$ which is difference between field frequency and atomic transition frequency. In various studies, the effects of detuning on the dynamics of atom-field interaction have been investigated. The Hamiltonian of the atom-field system in the rotating frame with $\Delta$ becomes\cite{Lambropoulos2007}

\begin{equation}
\hat{H}_{\text{det}} = \Delta \hat{\sigma}_{-}^{\text{A}} \hat{\sigma}_{+}^{\text{A}} + \lambda \left(\sqrt{\hat{N}_{\text{a}}}\,\hat{a}^{\dagger} \hat{\sigma}_{-}^{\text{A}} + \hat{a}\sqrt{\hat{N}_{\text{a}}}\, \hat{\sigma}_{+}^{\text{A}}\right)
+ \Delta \hat{\sigma}_{-}^{\text{B}} \hat{\sigma}_{+}^{\text{B}} + \lambda \left(\sqrt{\hat{N}_{\text{b}}}\,\hat{b}^{\dagger} \hat{\sigma}_{-}^{\text{B}} + \hat{b}\sqrt{\hat{N}_{\text{b}}}\, \hat{\sigma}_{+}^{\text{B}}\right),
\end{equation}
\begin{equation}
\hat{H'}_{\text{det}} = \Delta \hat{\sigma}_{-}^{\text{A}} \hat{\sigma}_{+}^{\text{A}} + \lambda (\hat{a}^{\dagger} \hat{\sigma}_{-}^{\text{A}} + \hat{a} \hat{\sigma}_{+}^{\text{A}}) + \Delta \hat{\sigma}_{-}^{\text{B}} \hat{\sigma}_{+}^{\text{B}} + \lambda (\hat{b}^{\dagger} \hat{\sigma}_{-}^{\text{B}} + \hat{b} \hat{\sigma}_{+}^{\text{B}}).
\end{equation}
To investigate the influence of detuning on the entanglement dynamics, we consider the average photon numbers $\bar{n}_c = 2$, $\bar{n}_s = 1.0$, and $\bar{n}_{th} = 0.1$, while varying the detuning parameter as $\Delta = 2, 5, 10$. The corresponding entanglement dynamics for the squeezed coherent thermal state (SCTS) in the intensity-dependent double Jaynes–Cummings model (IDDJCM) are illustrated in Fig.~\ref{fig_13}.

From the plots, it is evident that detuning $\Delta$ plays a crucial role in shaping the entanglement behavior and suppressing entanglement sudden deaths (ESDs). For a moderate detuning value $\Delta = 2$, the atom–atom entanglement $C(t)$ (Fig.~\ref{fig_13}(a), blue curve) shows a significant reduction in the duration of ESDs due to the emergence of additional smaller revival peaks. However, while these smaller revivals reduce ESD duration, the overall peak heights diminish with time, and the dynamics exhibit partial loss of periodicity.

In the case of atom A–field a entanglement $N(t)$, detuning destroys the periodic pulse structure entirely and enhances the amplitude, effectively eliminating ESDs. A similar behavior is observed for the atom A–field b entanglement, where $\Delta$ breaks the periodicity and shortens the ESD intervals, though not entirely removing them. For the field a–field b subsystem, $\Delta = 2$ successfully removes all ESDs, although the resulting dynamics become less regular and the periodic structure becomes distorted.

As detuning increases to $\Delta = 5$, the peak height of $C(t)$ further diminishes, while irregular smaller peaks appear sporadically, leading to inconsistent ESD durations across the time evolution. In this case, atom A–field a entanglement remains relatively unaffected, whereas atom A–field b entanglement experiences increased amplitude and partial removal of ESDs. For field–field entanglement, $N(t)$ loses its periodic nature entirely and its amplitude is significantly reduced.

For large detuning, $\Delta = 10$, the system displays more prominent oscillatory suppression. The atom–atom entanglement initially shows longer ESDs due to the disappearance of revival peaks. After a short revival, ESDs reappear, indicating that large detuning does not effectively prevent ESD in the long run. These findings are consistent with earlier results obtained for squeezed coherent and Glauber–Lachs states~\cite{Mandal_2024}, where detuning enhances atom–atom entanglement but limits energy and information exchange across subsystems.

The effect of detuning on SCTS within the standard DJCM framework is shown in Fig.~\ref{fig_14}. For $\Delta = 2$, the initial ESD in $C(t)$ is removed, yet the subsequent peak is suppressed, causing the next ESD to lengthen significantly (compare blue curves in Fig.~\ref{fig_2}(a) and Fig.~\ref{fig_14}(a)). For all atom–field and field–field subsystems, $N(t)$ decreases notably, indicating reduced entanglement sharing across the subsystems.

At higher detuning values $\Delta = 5$ and $10$, the amplitude of $C(t)$ initially increases but the corresponding ESDs become much longer, highlighting an accumulation of entanglement in the atom–atom subsystem at the expense of the others. Meanwhile, the atom–field and field–field entanglements decay further, with ESD appearing early in the field–field entanglement dynamics. This behavior suggests that increasing detuning transfers entanglement to certain subsystems (particularly atom–atom), from other subsystems. This is consistent with the previous results in \cite{Mandal_2024}. As a result, detuning serves as a control knob for selectively enhancing or suppressing subsystem entanglements, which is valuable in designing robust quantum information protocols.

\section{Conclusion}

In this work, we have systematically investigated the entanglement dynamics of atom--atom, atom--field, and field--field subsystems in the presence of squeezed coherent thermal states (SCTS) within both the intensity-dependent double Jaynes--Cummings model (IDDJCM) and the standard double Jaynes--Cummings model (DJCM). Our analysis reveals distinct behaviors in entanglement evolution depending on the model type and the nature of interactions introduced. In the IDDJCM, atom--atom and field--field entanglements remain largely unaffected by variations in the average number of squeezed photons when coherent and thermal photon numbers are held constant, whereas atom--field entanglements show noticeable sensitivity. In contrast, DJCM dynamics respond more strongly to increased squeezing, with entanglement sudden death (ESD) becoming more pronounced in the atom--atom and atom A--field b subsystems.

The inclusion of photon exchange interaction between the cavities plays a crucial role in removing ESDs from the dynamics, particularly in the atom--atom and field--field subsystems, for sufficiently large values of the interaction parameter \(\kappa\), in both DJCM and IDDJCM. However, this comes at the cost of diminishing atom--field entanglements. Notably, in DJCM, the field--field entanglement exhibits wave-packet-like structures under strong photon hopping, which are absent in IDDJCM. Similarly, the introduction of dipole--dipole interaction (\(g_d\)) effectively eliminates ESDs from the atom--atom and field--field subsystems in both models, although no wave packet formation is observed in this case. On the other hand, Ising-type interaction exhibits limited utility: while it can remove ESDs from field--field entanglement in IDDJCM, it fails to do so for atom--atom and atom--field subsystems, and has minimal effect on DJCM dynamics.

Kerr nonlinearity further enriches the entanglement landscape. In both models, it enhances \(C(t)\) and \(N(t)\) for several subsystems and efficiently removes ESDs, particularly in IDDJCM. However, for DJCM, field--field entanglement begins to decline beyond a critical Kerr strength \(\chi\), indicating a nonlinear redistribution of entanglement. Detuning also contributes to controlling entanglement dynamics by suppressing ESDs and promoting robustness. Increasing the detuning parameter \(\Delta\) leads to a rise in atom--atom entanglement while attenuating the strength of atom--field and field--field correlations, suggesting a transfer of entanglement into the atom--atom channel.

From an application perspective, the study of SCTS in the context of the IDDJCM offers valuable insights for quantum information science and quantum technology. The photon-number-dependent coupling in IDDJCM models nonlinear atom--field interactions relevant to real-world platforms such as superconducting qubits, trapped ions, and cavity QED systems. The mixed-state character of SCTS captures realistic environmental effects by combining coherence, squeezing, and thermal noise. These results hold particular relevance for the development of robust quantum communication protocols, fault-tolerant quantum gates, and long-lived quantum memories. By understanding how entanglement behaves under various interactions and noisy conditions, this study contributes to the ongoing effort to design quantum systems with tunable, stable, and controllable entanglement---essential ingredients for scalable quantum technologies.

\section*{Acknowledgements}
The author would like to thank Prof. M. V. Satyanarayana for all the valuable discussions and suggestions.

\bibliographystyle{naturemag}
\bibliography{scts_iddjcm_ref.bib}
\pagebreak
\appendix
\renewcommand{\thesection}{Appendix~\Alph{section}}

\section{Photon counting distribution for SCTS }\label{app_A}

The analytic expression for the PCD of SCTS can be written as\cite{PhysRevA.47.4474, PhysRevA.47.4487}

\begin{align}
P(l) =& \bra{l}\hat{\rho}_{\text{SCT}}\ket{l}\\
 =&~ \pi Q(0) \tilde{X}^{l}\sum_{q=0}^{l}\frac{1}{q!}\left(\frac{l}{q}\right)\Big|\frac{|\tilde{Y}|}{2 \tilde{X}}\Big|^{q}\nonumber\\
&\times \big|H_{q}((2Y)^{-1\slash 2} \tilde{Z})\big|^{2},
\end{align}
where $\pi Q(0) = R(0,0)$; $R$ is Glauber's $R$-function\cite{PhysRev.131.2766};

\begin{equation}
R(0,0) = \left[(1 + X)^{2} - |Y|^{2}\right]^{-1\slash 2}\exp\left\{- \frac{(1+X)|Z|^{2} + \frac{1}{2}[Y(Z^{*})^{2} + Y^{*}Z^{2}]}{(1+X)^{2}-|Y|^{2}}\right\},
\end{equation}
where
\begin{align}
X &= \bar{n}_{th} + (2 \bar{n}_{th} + 1) (\sinh r)^{2}, \\
Y &= - (2 \bar{n}_{th} + 1)e^{i\varphi} \sinh r \cosh r,\\
Z &= \alpha  \hspace{0.5cm} \text{(for SCTS)},\\
\end{align}
and
\begin{align}
\tilde{X} &= \frac{X(1 + X) - |Y|^{2}}{(1 + X)^{2} - |Y|^{2}}, \\
\tilde{Y} &= \frac{Y}{(1 + X)^{2} - |Y|^{2}},\\
\tilde{Z} &= \frac{(1 + X)Z + YZ^{*}}{(1 + X)^{2} - |Y|^{2}}.
\end{align}
If we write $\tilde{X}$, $\tilde{Y}$ and $\tilde{Z}$ in terms of $\bar{n}_{th}$ and $r$, we get
\begin{align}
\tilde{X} &= \frac{\bar{n}_{th}(\bar{n}_{th} + 1)}{\bar{n}_{th}^{2} + (\bar{n}_{th} + \frac{1}{2})[1 + \cosh (2r)]},\\
\tilde{Y} &= -\frac{e^{i\varphi}(\bar{n}_{th} + \frac{1}{2})\sinh(2r)}{\bar{n}_{th}^{2} + (\bar{n}_{th} + \frac{1}{2})[1 + \cosh (2r)]},\\
\tilde{Z} &= \frac{Z[\frac{1}{2} + (\bar{n}_{th}+ \frac{1}{2})\cosh r] - Z^{*}e^{i \varphi}(\bar{n}_{th}+ \frac{1}{2}) \sinh 2r}{\bar{n}_{th}^{2} + (\bar{n}_{th} + \frac{1}{2})[1 + \cosh (2r)]}, \nonumber\\
\end{align}
and $H_{q}$ is the Hermite polynomial. It is defined as
\begin{equation}
H_{q}(x) = \sum_{j=0}^{\lfloor\frac{ q}{2}\rfloor}\frac{(-1)^{j}q!}{j!(q-2j)!}(2x)^{q-2j}.
\end{equation}
\pagebreak

\section{Density matrix formalism for non-interacting Hamiltonian for DJCM and IDDJCM}\label{app_B}

The total initial density operator for the system can be written as
\begin{equation}
    \hat{\rho}_{\text{tot}}(0) = \sum_{i,j} c_{ij} \, |i\rangle \langle j| \otimes \hat{\rho}^{\text{a}}_{\text{F}}(0) \otimes \hat{\rho}^{\text{b}}_{\text{F}}(0).
\end{equation}
After evolution under the unitary operator the time evolved density operator becomes
\begin{equation}
    \hat{\rho}_{\text{tot}}(t) = \hat{U}(t) \, \hat{\rho}_{\text{tot}}(0) \, \hat{U}^\dagger(t).
\end{equation}
If we consider a single term in the summation
\begin{equation}
    \hat{\rho}_{\text{tot}}^{ij}(0) = |i\rangle \langle j| \otimes \hat{\rho}^{\text{a}}_{\text{F}}(0) \otimes \hat{\rho}^{\text{b}}_{\text{F}}(0),
\end{equation}
its time evolution can be expressed as
\begin{equation}
    \hat{\rho}_{\text{tot}}^{ij}(t) = \hat{U}(t) \left[ |i\rangle \langle j| \otimes \hat{\rho}^{\text{a}}_{\text{F}}(0) \otimes \hat{\rho}^{\text{b}}_{\text{F}}(0) \right] \hat{U}^\dagger(t).
\end{equation}
In the double Jaynes--Cummings model without cross-interactions, the total unitary operator can be factorized as
\begin{equation}
    \hat{U}(t) = \hat{U}^{(1)}(t) \otimes \hat{U}^{(2)}(t).
\end{equation}
where $\hat{U}^{(1)}(t)$ acts on atom A and field mode in cavity a and $\hat{U}^{(2)}(t)$ acts on atom B and field mode in cavity b.\\

\noindent\textbf{Time evolution for the term} $|e,g\rangle \langle e,g|$:\\

For $|i\rangle = |e,g\rangle$ and $\langle j| = \langle e,g|$:
\begin{equation}
    \hat{\rho}_{\text{tot}}^{eg,eg}(0) = \left[ |e\rangle \langle e| \otimes |g\rangle \langle g| \right] \otimes \hat{\rho}^{\text{a}}_{\text{F}}(0) \otimes \hat{\rho}^{\text{b}}_{\text{F}}(0).
\end{equation}
The time evolution becomes:
\begin{equation}
    \hat{\rho}_{\text{tot}}^{eg,eg}(t) = \left[ \hat{U}^{(1)}(t) |e\rangle \langle e|  \otimes\hat{\rho}^{\text{a}}_{\text{F}}(0) \hat{U}^{(1) \dagger}(t) \right]
     \otimes \left[ \hat{U}^{(2)}(t) |g\rangle \langle g| \otimes \hat{\rho}^{\text{b}}_{\text{F}}(0) \hat{U}^{(2) \dagger}(t) \right].
\end{equation}
$\hat{U}^{(1)}(t)$ and $\hat{U}^{(2)}(t)$ for DJCM can be expanded in the two dimensional subspace as \cite{gerry2005introductory}
\begin{equation}
\hat{U}^{(1, 2)}(t)= \begin{pmatrix} \cos(\lambda t\sqrt{(\hat{N} + 1 )}) &- i \frac{\hat{a}^{\dagger}\sin(\lambda t \sqrt{\hat{N} +1 })}{\sqrt{\hat{N}}}\\ - i \frac{\hat{a}\sin(\lambda t\sqrt{\hat{N}})}{\sqrt{\hat{N}}} & \cos(\lambda t\sqrt{\hat{N})}
\end{pmatrix}.
\end{equation}
For the intensity-dependent double Jaynes-Cummings model these unitary matrices can be expressed as 
\begin{equation}
\hat{U}^{(1, 2)}(t)= \begin{pmatrix} \cos(\lambda t (\hat{N} + 1 )) &- i \frac{\hat{a} \sin(\lambda t \hat{N})}{\sqrt{\hat{N}}}\\ - i \frac{\sin(\lambda t \hat{N}) \hat{a}^{\dagger}}{\sqrt{\hat{N}}} & \cos(\lambda t\hat{N})
\end{pmatrix}.
\end{equation}
We can write the unitary operator in more compact form as
\begin{equation}
\hat{U}^{(1, 2)}(t)= \begin{pmatrix} \hat{\mathscr{C}}(t) & \hat{\mathscr{S}}^{'}(t)\\ \hat{\mathscr{S}}(t) & \hat{\mathscr{C}}^{'}(t) \end{pmatrix},
\end{equation}
where
\begin{align}
\text{for DJCM:} \qquad
& \hat{\mathscr{C}}(t) = \cos(\lambda t \sqrt{\hat{N} + 1}), \qquad &&
\hat{\mathscr{C}}^{'}(t) = \cos(\lambda t \sqrt{\hat{N}} ), \\
& \hat{\mathscr{S}}(t) = -i \frac{\hat{a}^{\dagger} \sin(\lambda t \sqrt{\hat{N} + 1} )}{\sqrt{\hat{N}}}, \qquad &&
\hat{\mathscr{S}}^{'}(t) = -i \frac{\hat{a} \sin(\lambda t \sqrt{\hat{N}})}{\sqrt{\hat{N}}}, \\\nonumber
\\ 
\text{for IDDJCM:} \qquad
& \hat{\mathscr{C}}(t) = \cos\left(\lambda t (\hat{N} + 1) \right), \qquad &&
\hat{\mathscr{C}}^{'}(t) = \cos\left(\lambda t \hat{N} \right), \\
& \hat{\mathscr{S}}(t) =-i \frac{\sin\left(\lambda t \hat{N} \right) \hat{a}^{\dagger}}{\sqrt{\hat{N}}},\qquad &&
\hat{\mathscr{S}}^{'}(t) =  -i \frac{\hat{a} \sin\left(\lambda t \hat{N} \right)}{\sqrt{\hat{N}}}.
\end{align}
are the matrix elements of the unitary matrix. The Hermitian conjugate of $\hat{U}^{(1, 2)}(t)$ is 
 \begin{equation}
 \hat{U}^{(1, 2)\dagger}(t)=\begin{pmatrix} \hat{\mathscr{C}}(t) & -\hat{\mathscr{S}}^{'}(t)\\ -\hat{\mathscr{S}}(t) & \hat{\mathscr{C}}^{'}(t) \end{pmatrix}.
 \end{equation}

\noindent \textbf{To calculate the term:} $\hat{U}^{(1)}(t) (|e\rangle \langle e|  \otimes\hat{\rho}^{\text{a}}_{\text{F}}(0)) \hat{U}^{(1) \dagger}(t)$:\\

In matrix form
\begin{equation}
\ket{e}\bra{e}=\begin{pmatrix} 1 & 0 \\ 0 & 0 \end{pmatrix}.
\end{equation}
Under the time evolution 
\begin{equation}
\hat{U}^{(1)}(t) \left(|e\rangle \langle e|  \otimes\hat{\rho}^{\text{a}}_{\text{F}}(0)\right) \hat{U}^{(1) \dagger}(t)= \begin{pmatrix} \hat{\mathscr{C}}(t)\hat{\rho}^{\text{a}}_{\text{F}}(0)\hat{\mathscr{C}}^{'}(t) & -\hat{\mathscr{C}}(t)\hat{\rho}^{\text{a}}_{\text{F}}(0)\hat{\mathscr{S}}^{'}(t) \\ \hat{\mathscr{S}}(t)\hat{\rho}^{\text{a}}_{\text{F}}(0)\hat{\mathscr{C}}^{'}(t) & - \hat{\mathscr{S}}(t)\hat{\rho}^{\text{a}}_{\text{F}}(0)\hat{\mathscr{S}}^{'}(t) \end{pmatrix}.
\label{rho_mat}
\end{equation}
Similarly,
\begin{equation}  
\hat{U}^{(2)}(t) \left(|g\rangle \langle g| \otimes \hat{\rho}^{\text{b}}_{\text{F}}(0)\right) \hat{U}^{(2) \dagger}(t)= \begin{pmatrix} -\hat{\mathscr{S}^{'}}(t)\hat{\rho}^{\text{b}}_{\text{F}}(0)\hat{\mathscr{S}}(t) & \hat{\mathscr{S}^{'}}(t)\hat{\rho}^{\text{b}}_{\text{F}}(0)\hat{\mathscr{C}}(t) \\ -\hat{\mathscr{C}}(t)\hat{\rho}^{\text{b}}_{\text{F}}(0)\hat{\mathscr{S}}(t) &  \hat{\mathscr{C}}(t)\hat{\rho}^{\text{b}}_{\text{F}}(0)\hat{\mathscr{C}}^{'}(t) \end{pmatrix}.
\end{equation}
So, the matrix form of $\hat{\rho}^{eg,eg}(t)$ is \\
\begin{equation*}
\begin{pmatrix} \hat{\mathscr{C}}(t)\hat{\rho}^{\text{a}}_{\text{F}}(0)\hat{\mathscr{C}}^{'}(t) & -\hat{\mathscr{C}}(t)\hat{\rho}^{\text{a}}_{\text{F}}(0)\hat{\mathscr{S}}^{'}(t) \\ \hat{\mathscr{S}}(t)\hat{\rho}^{\text{a}}_{\text{F}}(0)\hat{\mathscr{C}}^{'}(t) & - \hat{\mathscr{S}}(t)\hat{\rho}^{\text{a}}_{\text{F}}(0)\hat{\mathscr{S}}^{'}(t) \end{pmatrix}\otimes 
\begin{pmatrix} -\hat{\mathscr{S}^{'}}(t)\hat{\rho}^{\text{b}}_{\text{F}}(0)\hat{\mathscr{S}}(t) & \hat{\mathscr{S}^{'}}(t)\hat{\rho}^{\text{b}}_{\text{F}}(0)\hat{\mathscr{C}}(t) \\ -\hat{\mathscr{C}}(t)\hat{\rho}^{\text{b}}_{\text{F}}(0)\hat{\mathscr{S}}(t) &  \hat{\mathscr{C}}(t)\hat{\rho}^{\text{b}}_{\text{F}}(0)\hat{\mathscr{C}}^{'}(t) \end{pmatrix}.
\end{equation*}
which after calculating the tensor product becomes
\[
\resizebox{0.75\textwidth}{!}{%
\renewcommand{\arraystretch}{1.5}%
$
\begin{aligned}
&\hspace{-2cm} 
\left[
\begin{array}{cc}
-\hat{\mathscr{C}}(t)\hat{\rho}^{\text{a}}_{\text{F}}(0)\hat{\mathscr{C}}'(t)\cdot \hat{\mathscr{S}}'(t)\hat{\rho}^{\text{b}}_{\text{F}}(0)\hat{\mathscr{S}}(t) &
\hat{\mathscr{C}}(t)\hat{\rho}^{\text{a}}_{\text{F}}(0)\hat{\mathscr{C}}'(t) \cdot \hat{\mathscr{S}}'(t)\hat{\rho}^{\text{b}}_{\text{F}}(0)\hat{\mathscr{C}}(t) \\
-\hat{\mathscr{C}}(t)\hat{\rho}^{\text{a}}_{\text{F}}(0)\hat{\mathscr{C}}'(t)\cdot \hat{\mathscr{C}}(t)\hat{\rho}^{\text{b}}_{\text{F}}(0)\hat{\mathscr{S}}(t) &
\hat{\mathscr{C}}(t)\hat{\rho}^{\text{a}}_{\text{F}}(0)\hat{\mathscr{C}}'(t)\cdot \hat{\mathscr{C}}(t)\hat{\rho}^{\text{b}}_{\text{F}}(0)\hat{\mathscr{C}}'(t) \\
\hat{\mathscr{S}}(t)\hat{\rho}^{\text{a}}_{\text{F}}(0)\hat{\mathscr{C}}'(t)\cdot \hat{\mathscr{S}}'(t)\hat{\rho}^{\text{b}}_{\text{F}}(0)\hat{\mathscr{S}}(t) &
-\hat{\mathscr{S}}(t)\hat{\rho}^{\text{a}}_{\text{F}}(0)\hat{\mathscr{S}}'(t) \cdot \hat{\mathscr{S}}'(t)\hat{\rho}^{\text{b}}_{\text{F}}(0)\hat{\mathscr{C}}(t) \\
-\hat{\mathscr{S}}(t)\hat{\rho}^{\text{a}}_{\text{F}}(0)\hat{\mathscr{C}}'(t)\cdot \hat{\mathscr{C}}(t)\hat{\rho}^{\text{b}}_{\text{F}}(0)\hat{\mathscr{S}}(t) &
\hat{\mathscr{S}}(t)\hat{\rho}^{\text{a}}_{\text{F}}(0)\hat{\mathscr{C}}'(t) \cdot \hat{\mathscr{C}}(t)\hat{\rho}^{\text{b}}_{\text{F}}(0)\hat{\mathscr{C}}'(t)
\end{array}
\right. \\[2ex]
&\hspace{2cm} 
\left.
\begin{array}{cc}
-\hat{\mathscr{C}}(t)\hat{\rho}^{\text{a}}_{\text{F}}(0)\hat{\mathscr{S}}'(t) \cdot \hat{\mathscr{S}}'(t)\hat{\rho}^{\text{b}}_{\text{F}}(0)\hat{\mathscr{S}}(t) &
\hat{\mathscr{C}}(t)\hat{\rho}^{\text{a}}_{\text{F}}(0)\hat{\mathscr{S}}'(t) \cdot \hat{\mathscr{S}}'(t)\hat{\rho}^{\text{b}}_{\text{F}}(0)\hat{\mathscr{C}}(t) \\
-\hat{\mathscr{C}}(t)\hat{\rho}^{\text{a}}_{\text{F}}(0)\hat{\mathscr{S}}'(t) \cdot \hat{\mathscr{C}}(t)\hat{\rho}^{\text{b}}_{\text{F}}(0)\hat{\mathscr{S}}(t) &
-\hat{\mathscr{C}}(t)\hat{\rho}^{\text{a}}_{\text{F}}(0)\hat{\mathscr{S}}'(t) \cdot \hat{\mathscr{C}}(t)\hat{\rho}^{\text{b}}_{\text{F}}(0)\hat{\mathscr{C}}'(t) \\
\hat{\mathscr{S}}(t)\hat{\rho}^{\text{a}}_{\text{F}}(0)\hat{\mathscr{S}}'(t) \cdot \hat{\mathscr{S}}'(t)\hat{\rho}^{\text{b}}_{\text{F}}(0)\hat{\mathscr{S}}(t) &
\hat{\mathscr{S}}(t)\hat{\rho}^{\text{a}}_{\text{F}}(0)\hat{\mathscr{C}}'(t) \cdot \hat{\mathscr{S}}'(t)\hat{\rho}^{\text{b}}_{\text{F}}(0)\hat{\mathscr{C}}(t) \\
\hat{\mathscr{S}}(t)\hat{\rho}^{\text{a}}_{\text{F}}(0)\hat{\mathscr{S}}'(t) \cdot \hat{\mathscr{C}}(t)\hat{\rho}^{\text{b}}_{\text{F}}(0)\hat{\mathscr{S}}(t) &
-\hat{\mathscr{S}}(t)\hat{\rho}^{\text{a}}_{\text{F}}(0)\hat{\mathscr{S}}'(t) \cdot \hat{\mathscr{C}}(t)\hat{\rho}^{\text{b}}_{\text{F}}(0)\hat{\mathscr{C}}'(t)
\end{array}
\right].
\end{aligned}
$
}
\]
Similarly it can be calculated for other terms. For example, the cross-term $|e,g\rangle \langle g,e|$

\begin{equation}
    \hat{\rho}^{eg,ge}(0) = \left[ |e\rangle \langle g| \otimes |g\rangle \langle e| \right]  \hat{\rho}^{\text{a}}_{\text{F}}(0) \otimes \hat{\rho}^{\text{b}}_{\text{F}}(0).
\end{equation}
The time evolution becomes:
\begin{equation}
    \hat{\rho}^{eg,ge}(t) =  \left[ \hat{U}^{(1)}(t) (|e\rangle \langle g| \otimes\hat{\rho}^{\text{a}}_{\text{F}}(0) \hat{U}^{(1) \dagger}(t) \right] 
    \otimes \left[ (\hat{U}^{(2)}(t) |g\rangle \langle e|\otimes \hat{\rho}^{\text{b}}_{\text{F}}(0) \hat{U}^{(2) \dagger}(t) \right].
\end{equation}
The final density matrix in the atomic basis would be $4 \times 4$ matrix. However, if we express the density operator for the field as a square matrix of dimension $n\times n$, the total matrix for the whole system becomes the size of $4n^2 \times 4n^2$.

\pagebreak

\section{Additional plots of entanglement for IDDJCM and DJCM}\label{app_C}

\begin{figure}[ht!]
    \centering
    \includegraphics[scale = 0.45]{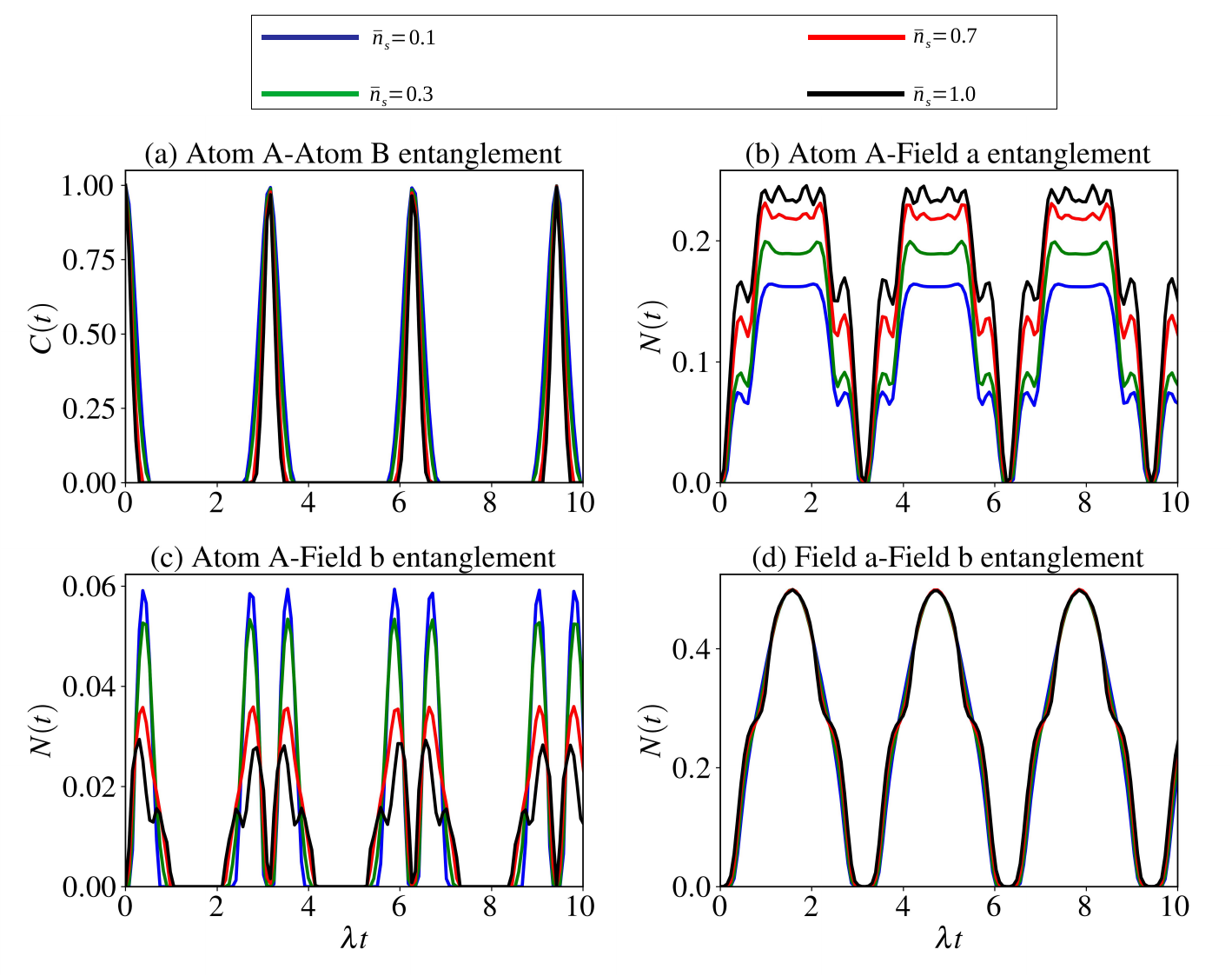}
    \caption{Entanglement dynamics for atom A-atom B, atom A-field a, atom a-field b and field a-field b with atoms in a Bell state and field in SCTS for IDDJCM. The values of the parameters used in these plots are $\bar{n}_c = 2$, $\bar{n}_{th} = 0.1$, $\bar{n}_s = 0.1, 0.3, 0.5, 1.0$ and $\theta = \frac{\pi}{4}$.}
    \label{fig_15}
\end{figure}

\begin{figure}[ht!]
    \centering
    \includegraphics[scale = 0.35]{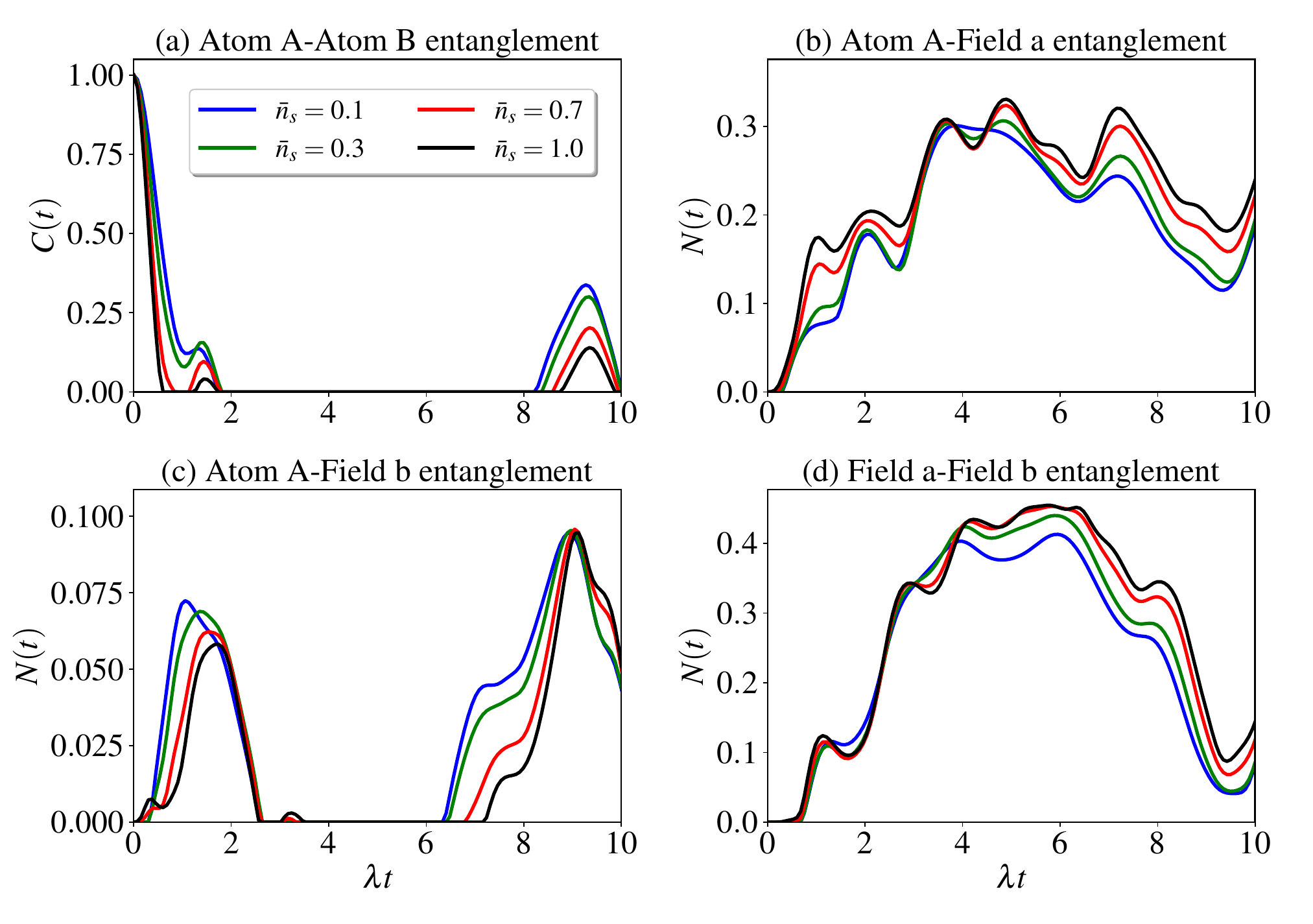}
    \caption{Entanglement dynamics for atom A-atom B, atom A-field a, atom a-field b and field a-field b with atoms in a Bell state and field in SCTS for DJCM. The values of the parameters used in these plots are $\bar{n}_c = 2$, $\bar{n}_{th} = 0.1$, $\bar{n}_s = 0.1, 0.3, 0.5, 1.0$ and $\theta = \frac{\pi}{4}$.}
    \label{fig_16}
\end{figure}

Figures~\ref{fig_15} and~\ref{fig_16} illustrate the time evolution of entanglement for all subsystems, presented together in each plot to provide a comparative perspective. The periodic nature of entanglement dynamics in the IDDJCM is clearly visible in Fig.~\ref{fig_15}, while the non-periodic behavior characteristic of the DJCM is distinctly observed in Fig.~\ref{fig_16}.

\pagebreak

\section{Effects of thermal photons on concurrence}\label{app_D}

\begin{figure}[ht!]
    \centering
    \includegraphics[scale = 0.6]{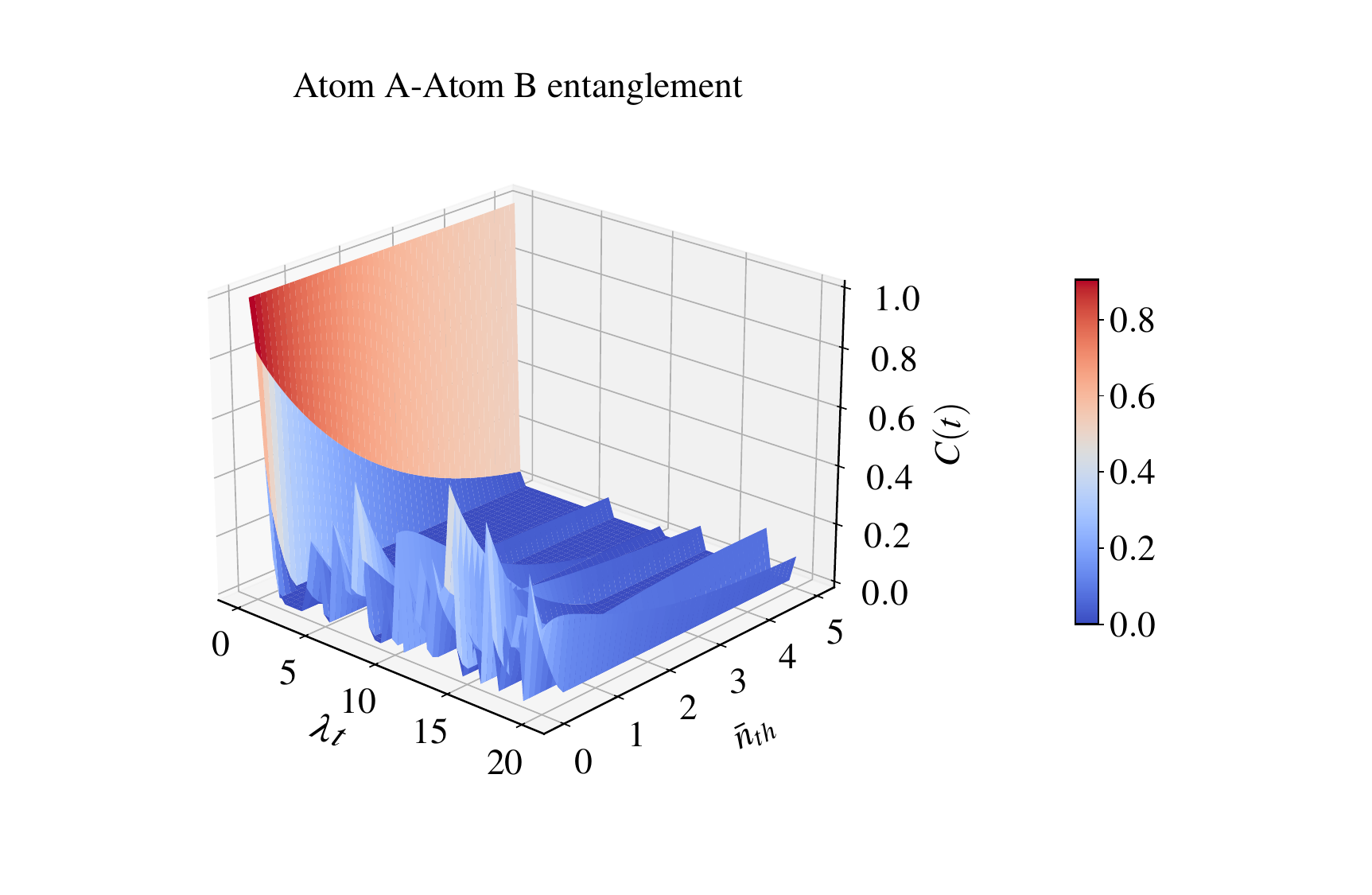}
    \caption{3D plot of $C(t)$ \textit{versus} $\bar{n}_{th}$ and $\lambda t$ for intensity-dependent double Jaynes-Cummings model. \\
    Here,  $\bar{n}_c = 2, \bar{n}_s = 0$ and $\bar{n}_{th}$ is varied from $0$ to $5$.}
    \label{ct_nth}
\end{figure}

We present the 3D graphs of $C(t)$ with respect too varying thermal photons in Fig. \ref{ct_nth}. Here, $\bar{n}_c = 2$ and $\bar{n}_{th}, \bar{n}_s$ are varied from $0$ to $5$. With these choices of the parameters, two cases can be compared (i) when $\bar{n}_c < \bar{n}_{th}$ and (ii)  when $\bar{n}_c > \bar{n}_{th}$. From the figure, it can be observed that $C(t)$ starts from maximum value for each value of $\bar{n}_{th}$; however, for $\bar{n}_c > \bar{n}_{th}$, large number of peaks appear in the dynamics and the length of ESDs are smaller while $\bar{n}_c < \bar{n}_s$, many peaks disappear and amplitude also decreases. This makes the lengths of ESDs to increase in the dynamics. The atom-field entanglement $N(t)$ decreases significantly with the addition of thermal photons in the system.

\pagebreak

\section{Effects of Kerr-nonlinearity on entanglement (3D plot)}\label{app_E}
\begin{figure}[ht!]
    \centering
    \includegraphics[scale = 0.4]{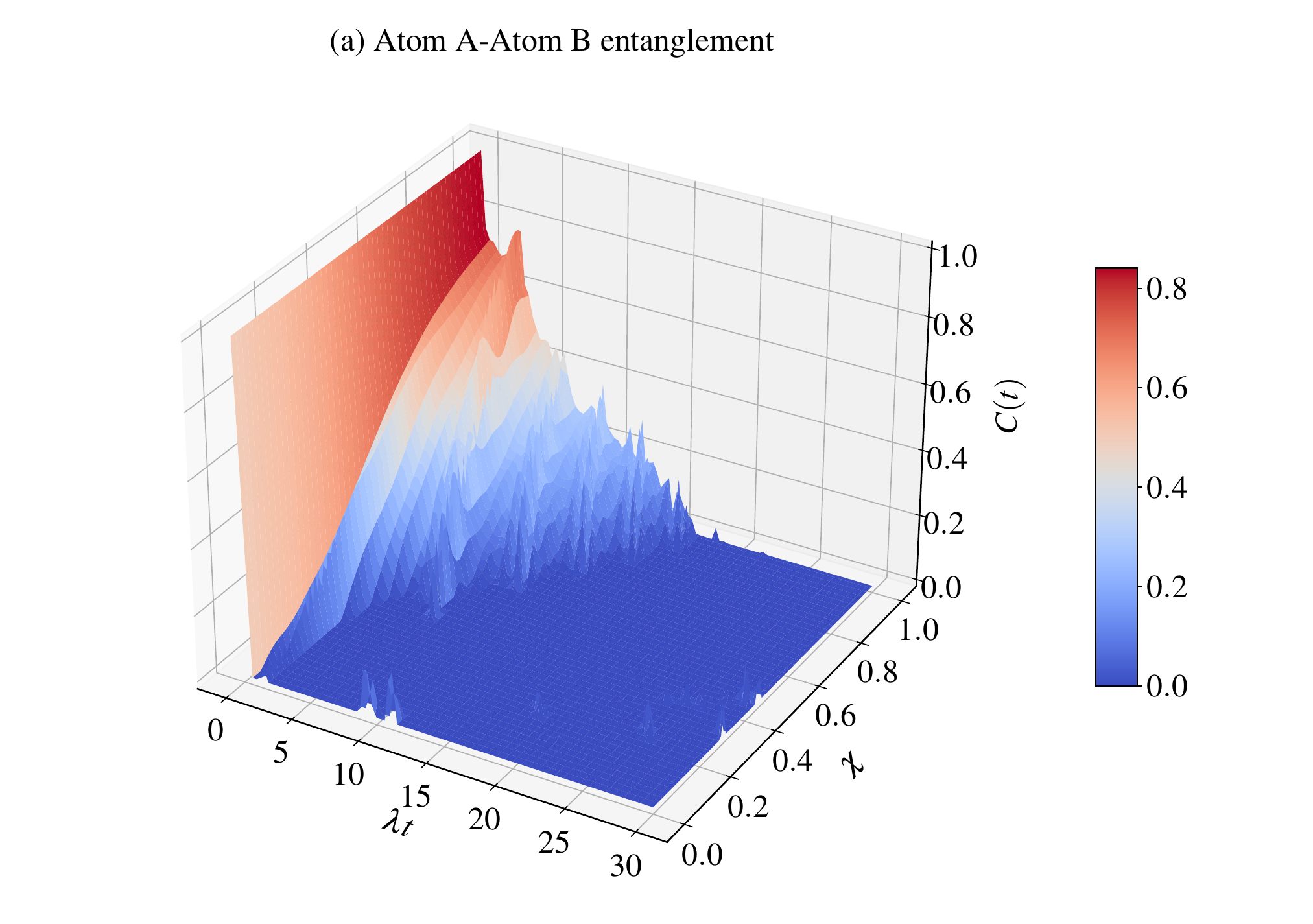}
    \caption{3D graphs of $C(t)$ \textit{versus} $\lambda t$ and $\chi$ for IDDJCM with Kerr-nonlinearity inside cavities with atoms in a Bell state.The values of the parameters used in these plots are $\bar{n}_c = 2$, $\bar{n}_{th} = 0.1$, $\bar{n}_s = 1.0$, $\chi = [0, 1]$ and $\theta = \frac{\pi}{4}$.} 
    \label{kerr_3d}
\end{figure}

Figure \ref{kerr_3d} shows how the concurrence of the atom-atom subystem evolves with time for varying $\chi$. Specifically, it is observed that for $\chi \approx 0.3$, the envelope of $C(t)$ reaches a relative minimum compared to neighboring values of $\chi$. The time window was also extended to confirm the long-time behavior. This plot confirms that there is a region around $\chi \approx 0.3$ where the concurrence temporarily decreases in amplitude, and then increases again for larger values of $\chi$.

\end{document}